\documentclass[review,sort&compress]{elsarticle}

\usepackage{lineno,hyperref}
\modulolinenumbers[5]
\usepackage{comment}

\usepackage{graphicx}
\graphicspath{{./Images/}}

\usepackage{amssymb}

\newcommand{\fg}[1]{\mbox{\pmb{$#1$}}}
\newcommand{\bey}{\begin{eqnarray}}
\newcommand{\eey}{\end{eqnarray}}
\newcommand{\vep}{\varepsilon}
\newcommand{\fvep}{\fg \varepsilon}
\newcommand{\fsg}{\fg \sigma}
\newcommand{\sg}{\sigma}
\newcommand{\bec}{\begin{center}}
\newcommand{\eec}{\end{center}}
\newcommand{\bs}[1]{\mbox{\pmb{$#1$}}}
\newcommand{\drop}[1]{}

\usepackage{amsbsy}
\usepackage{amsmath}
\usepackage{amsfonts}
\usepackage{amssymb}
\usepackage{textcomp}
\usepackage{cleveref}
\usepackage[margin=2.5cm]{geometry}
\usepackage{xcolor}
\usepackage{rotating}
\usepackage{float}

\begin{document}
	
	\begin{frontmatter}
		
		
		\title{ Simulations of multivariant Si I to Si II phase transformation in polycrystalline silicon with finite-strain scale-free phase-field approach}
		

		
	\author[1]{Hamed Babaei}
 \author[1]{Raghunandan Pratoori}
 		\author[1,2,3]{Valery~I. Levitas}
 		\address[1]{Department of Aerospace Engineering, Iowa State University, Ames, IA 50011, USA}
 		\address[2]{Departments of Mechanical Engineering, Iowa State University, Ames, IA 50011, USA}
 		\address[3]{Ames Laboratory, Division of Materials Science and Engineering, Ames, IA 50011, USA}

 		\begin{abstract}
Scale-free phase-field approach (PFA) at large strains and corresponding finite element method (FEM) simulations for multivariant martensitic phase transformation (PT) from cubic Si I to tetragonal Si II in a polycrystalline aggregate are presented. Important features of the model are large and very anisotropic transformation strain tensor $\fvep_{t}=\{0.1753;0.1753; -0.447\}$ and stress-tensor dependent athermal dissipative threshold for PT, which produce essential challenges for computations. 3D polycrystals with 55 and 910 stochastically oriented grains are subjected to uniaxial strain- and stress-controlled loadings under periodic boundary conditions and zero averaged lateral strains. Coupled evolution of discrete martensitic microstructure, volume fractions of martensitic variants and Si II, stress and transformation strain tensors, and texture are presented and analyzed. Macroscopic variables effectively representing multivariant transformational behavior are introduced. Macroscopic stress-strain and transformational behavior for 55 and 910 grains are close (less than 10\% difference). This allows the determination of macroscopic constitutive equations by treating aggregate with a small number of grains. Large transformation strains and grain boundaries lead to huge internal stresses of tens GPa, which affect microstructure evolution and macroscopic behavior. In contrast to a single crystal, the local mechanical instabilities due to PT and negative local tangent modulus are stabilized at the macroscale by arresting/slowing the growth of Si II regions by the grain boundaries and generating the internal back stresses. This leads to increasing stress during PT. The developed methodology can be used for studying similar PTs with large transformation strains and for further development by including plastic strain and strain-induced PTs.
		\end{abstract}
		
		
	\end{frontmatter}

	\section{Introduction}
	Silicon is the second most abundant material in the earth's crust.
The semiconducting Si I phase (cubic diamond lattice, $Fd3m$ space group)  is extensively used in microelectronics, integrated circuits, photovoltaics, and MEMS/ NEMS technologies. Single-crystal Si is also used in high-power lasers. Polycrystalline Si is widely used in solar panels~\citep{heath2020research}, thin transistors~\citep{chelikowsky2004introduction}, and very large-scale integration (VLSI) manufacturing.
It also has low toxicity and high stability.
Due to high demands,  the recent  CHIPS and Science Act will provide new funding to boost the research and manufacturing of semiconductors in the US.
Under the pressure of 10-16 GPa, semiconducting  Si I transforms to metallic phase Si II ($\beta$-tin structure, $I4_1/amd$ space group).
Si I is very strong and brittle, and hence its  bulk hardness  is 12 GPa  and is determined by Si I$\rightarrow$Si II PT
rather then dislocational plasticity \citep{Domnich-2004,Kiran2015NanoindentationGermanium}.
Stresses exceed 10 GPa for machining (turning, polishing, scratching, etc.) of single and polycrystalline Si \citep{Goel2015DiamondSimulation}; such loads cause plastic flow, Si I$\rightarrow$Si II and some other  PTs, e.g., amorphization. High-pressure torsion of Si at 24 GPa is used to produce nanostructured metastable phases \citep{Ikoma2012PhaseTorsion,Ikoma2014FabricationTorsion}.
Machining of strong brittle semiconducting Si-I  is accompanied by microcrack propagation inside the bulk.
 PT from  Si I to ductile and weaker Si II is utilized to  develop  ductile machining regimes  \cite{patten2019ductile}, which reduces forces,  energy, and damage.
 This also may eliminate the necessity of using chemical additives during machining, which  brings definite environmental benefits by reducing pollution.

Since transformation strain tensor that describes the transformation of cubic to tetragonal Si I $\rightarrow$ Si II
PT has large and very anisotropic principal components
 $\fvep_{t}=\{0.1753;0.1753; -0.447\} $ \citep{malyushitskaya1999mechanisms,Levitas2017LatticeCriterion,Levitas2017Triaxial-Stress-InducedPhases}, it is clear from thermodynamics that deviatoric part of the stress tensor should
 strongly affect this PT. Both pressure- and stress-induced PTs start at pre-existing defects (different dislocation configurations, grain boundaries), which represent stress concentrators.
 However, in many of the  applications, like turning, polishing, scratching, friction,
high-pressure torsion and ball milling, PTs occur during plastic deformation.
According to classification \citep{Levitas-PRB-2004,levitas2018high,Levitas2019High-pressureAnvils}, such PTs are called plastic strain-induced PT under high pressure, and they occur at defects permanently generated during plastic flow.
Strain-induced PTs require completely different thermodynamic,
kinetic, and experimental treatments than pressure- and
stress-induced PTs.

There are numerous very strong effects of plastic deformations on PTs, summarized in
 \cite{Blank2013,Bridgman1935EffectsPressure,edalati2016review,Gao2019,Levitas-PRB-2004,levitas2018high,Levitas2019High-pressureAnvils};
one of the most important is a drastic   reduction in PT pressure. Thus,
plastic strain-induced PTs from graphite to hexagonal and cubic diamonds were obtained at 0.4 and 0.7 GPa,
 50 and 100 times lower than under hydrostatic loading, respectively, and well below the phase equilibrium pressure of 2.45 GPa \citep{Gao2019}. About an order of magnitude reduction in PT pressure was reported for PT from
 rhombohedral to cubic BN  \citep{Levitas2002Low-pressureTheory}, hexagonal to wurtzitic BN  \citep{ji2012shear}, and
 from $\alpha$ to $\omega$ Zr \citep{Pandey-Levitas-2020,levitas2022laws}.

The effect of plastic straining on PTs in Si is also very strong but more sophisticated.
Thus, under compression and shear in rotational diamond anvil cell
 \citep{aleksandrova1993phase}, Si II was obtained at 5.2 GPa, but not directly from Si I, but via
 Si III. However, in these experiments,
  only optical, pressure, and electric resistivity measurements were utilized without in situ x-ray diffraction.
  In our recent in situ x-ray diffraction experiments \citep{Sorb2022},
  the PTs pressure for direct Si I$\rightarrow$Si II PT was
   reduced from 13.5 (hydrostatic loading) to 2.5 GPa (plastic straining) for micron size Si particles and from 16.2 GPa to 0.4 GPa for 100 nm Si nanoparticles, i.e., by a factor of 40.5 (and 26.3 below the phase equilibrium pressure of 10.5 GPa \citep{Voronin-etal-2003}).

To understand the effect of stress tensor and plastic strain on Si I$\rightarrow$Si II PT, various techniques at multiple scales are used.
With first principle simulations, the lattice instability for Si I under  two-parametric   loadings
was studied in
\cite{Pokluda-etal-2015,Umeno-Cerny-PRB-08,Telyatniketal-16,Cernyetal-JPCM-13}. Although it is not specified,
it is related to Si I$\rightarrow$Si II PT.
The stress-strain behavior and elastic instabilities leading to Si I$\rightarrow$Si II PT under all six components of the  stress tensor were determined in \cite{zarkevich2018lattice,Chen-etal-NPJ-2020}. While this seems to be impossible due to a large number of combinations, the following solution was found. First, the analytical expression for
crystal lattice instability criterion was formulated within the PFA   \citep{Levitas2013Phase-fieldStrains,Levitas2017LatticeCriterion,Levitas2017Triaxial-Stress-InducedPhases}.
Then it was confirmed and quantified with first principle simulations in \cite{zarkevich2018lattice}.

A similar approach was realized earlier with molecular dynamics simulations \citep{Levitas2017LatticeCriterion,Levitas2017Triaxial-Stress-InducedPhases}
under the action of three normal stresses. Molecular dynamics simulations of PTs in Si during various loadings, nanoindentation, scratching,
and surface processing were performed in \cite{valentini2007phase,chrobak2011deconfinement,Chen_2019,Zhang2019MolecularSilicon,Kiran2015NanoindentationGermanium,Chen2022NontrivialTransformation}.

Nanoscale PFA for PT Si I$\rightarrow$Si II in a single crystal with corresponding simulations was developed
in \cite{Levitas2018PhaseTheory,Babaei_2018,Babaei-Levitas-ActaMat-2019,Babaei2020Stress-MeasureTransformations,Babaei2019AlgorithmicStrains}. It was calibrated by results of molecular dynamics simulations in \cite{Levitas2017LatticeCriterion,Levitas2017Triaxial-Stress-InducedPhases}.
The effect of a single dislocation on this PT under uniaxial compression is modeled in  \cite{Babaei-Levitas-ActaMat-2019}.
The main general problem with nanoscale PFA  is that  the width of the martensitic interface is ~1 nm and one needs at least 3-5 finite elements within the interface \citep{Levitas2011Phase-fieldEnergy}, making problem  computationally expensive for large samples. Hence this model can be used to treat nano-sized samples only.

Here, we will consider scale-free PFA for modeling discrete martensitic microstructure. It was  developed for small strains in
	\cite{Levitas-etal-2004,Idesman-etal-2005} and updated and applied to NiTi shape memory alloy in  \cite{Esfahani-etal-2018}.   The only finite-strain generalization of the scale-free model for finite strains is presented in \cite{babaei2020finite}. This model was applied for simulations of Si I$\rightarrow$Si II PT in a single crystal.

The main differences between scale-free and nanoscale PFAs are:
	\begin{enumerate}
	\item The energy term that includes gradients of the order parameters  and determines the phase interface widths and the interface energies is excluded. This makes the model scale-free.  Interface width is getting equal to a single finite element, which is much more computationally economical than in the nanoscale approach, where 3-5 elements are usually required
to reproduce analytical solution for an interface \citep{Levitas2011Phase-fieldEnergy}.
However, this leads to two problems. (a) The solution is mesh dependent. However, detailed computational experiments in \cite{Levitas-etal-2004,Idesman-etal-2005,babaei2020finite} show that the solution becomes practically mesh-independent after the mesh size is 80 times smaller than the sample size.
(b) Since the volume fraction of martensite varies from 0 to 1 within the one-element thick interface, for large transformation strains, there are large strain gradients within the interface, which leads often to divergence in the FEM solution.

    \item The interfaces between individual martensitic variants are not resolved. Each of these variant  has a width of $ d\simeq10\;\text{nm} $, i.e., thousands of interfaces at the  microscale, making the problem computationally impractical. Here, martensite is considered a mixture of martensitic variants with corresponding volume fractions.
	    \item Since there is no need to reproduce atomic level energy landscape versus order parameters,
the linear mixture rule is applied for  all material properties. This is in contrast to higher-order  polynomials in order parameters for nanoscale PFA. Also, the athermal dissipative threshold for interface propagation (interface friction)
can be easily introduced in the scale-free model, while this is a problem for the nanoscale model.
  \item The  volume fraction of the martensite is the order parameter, i.e., it  is responsible for the material instability and transformation strain localization at the interface between austenite and martensite. The volume fractions of the individual martensitic variants are only internal variables and do not produce any instabilities.
      That is how we eliminate interfaces between martensitic variants.
	\end{enumerate}
All the above features allow us to economically model multivariant martensitic PTs in a sample of arbitrary size.

The next natural step is to simulate PTs in Si and discrete martensitic microstructure in a polycrystalline sample, which was not done yet. This is also an important step in studying plastic strain-induced PTs. The main hypothesis  \citep{Levitas-PRB-2004} is that they initiate
at the tip of dislocation pileups against grain boundaries, which causes strong stress concentration for all stress components proportional to the number of dislocations in a pileup. These stresses drastically reduce the required external pressure needed for
nucleation and further growth. Nanoscale PFA allows us to treat a bicrystal and to qualitatively prove this hypothesis
\citep{Levitas-Mahdi-Nanoscale-14,Javanbakht2018NanoscaleStudy,Javanbakht2016PhaseShear}, however, for 2D and small strain formulation. Our scale-free PFA \citep{Levitas2018Scale-FreeMicrostructure,Esfahani-etal-2020} has been applied to 2D polycrystalline samples
with several dozen grains and also for small strain formulation.

	In this work, we extend the modeling presented in \cite{babaei2020finite} to Si I$\rightarrow$Si II PT in 3D polycrystalline samples with up to 1000 grains.
The main challenge is to reach convergence of the computational procedure due to strong nonlinearities, large transformation strain localization with large gradients within a single-element diffuse interface, and potential elastic instabilities. For a single crystal, just two loadings, uniaxial and hydrostatic, have been considered in  \cite{babaei2020finite}. However, each grain is subjected to different complex and heterogeneous loading in a polycrystal.
While we used the simplest quadratic in elastic Lagrangian strain  expression (\ref{elastic_energy}) for the elastic energy, it is shown in \cite{Levitas_2017} that even simple uniaxial compression leads to elastic instability. This instability can cause additional strain localization and divergence of solution. Due to the variety of heterogeneous complex loadings, different for different grains, chances for various elastic instabilities and divergence are very high. That is why several computational parameters
have been varied to reach convergence of the solution. Periodic boundary conditions along the lateral surfaces
played important part in the avoiding divergence.
Another problem is to adequately quantitatively present the evolution of the local and average volume fraction of martensitic
 variants; a straightforward approach leads to contradictory results.

The paper is organized as follows.
In Section \ref{model} complete system of coupled PFA and nonlinear mechanics equations is presented.
Materials parameters for the model are given in Section  \ref{Model parameters}.
Problem formulations, results of simulations, and their analyses are presented in Section \ref{results}.
Concluding remarks are summarized in  Section \ref{Conclusion}.	
 Vectors and tensors are designated with boldface symbols. We designate contractions of tensors ${\fg A}=\{A_{ij}\}$ and ${\fg B}=\{B_{ji}\}$ over one and two indices as ${\fg A}{\fg \cdot}{\fg B}=\{A_{ij}\,B_{jk}\}$ and ${\fg A}{\fg :}{\fg B}=A_{ij}\,B_{ji}$. The transpose of ${\fg A}$ is ${\fg A}^{ T}$; ${\fg I}$ is the unit tensor; $\fg \nabla_0$ is the gradient operator in the undeformed state.
	
	\section{Model description}\label{model}

	Here, we expand  the microscale model for multivariant martensitic PTs developed in \cite{babaei2020finite}
for a polycrystalline elastic materials.
	
	
	\subsection{Kinematics:}
	
Let us consider polycrystalline Si I aggregate in the undeformed stress-free configuration $  \Omega_0  $.
The orientation  of each  grain is characterized by the rotation tensor ${\bs R_g}$ in the undeformed configuration,  which rotates local cubic crystallographic axes of the grain to the global coordinate system. Tensors ${\bs R_g}$ do not evolve during loading and PTs.
	The deformation gradient $\bs F$ is multiplicatively  split into the  elastic $\bs F_e$ and the transformational   $\bs F_t$ parts:
	\begin{equation}
		{\bs F} := \bs \nabla_0 {\bs r}= {\bs F}_e{\bs{\cdot}}{\bs F}_t;    \qquad {\bs F}_e={\bs R}_e\cdot{\bs U}_e;     \qquad {\bs F}_t={\bs F}_t^T.
		\label{Fdecom}
	\end{equation}
	Here, ${\bs r}$ is the position vector in the current deformed configuration $  \Omega  $; ${\bs F}_t$ is defined as ${\bs F}$ after complete local stress release (producing the  intermediate configuration $\Omega_t$) and is considered to be rotation-free; ${\bs U}_e$ is the symmetric elastic right stretch tensor and
${\bs R}_e$ is the orthogonal lattice rotation tensor during loading and PT, which determines texture evolution.
	The transformation deformation gradient  is defined using the mixture rule for $m$ variants as
	\begin{equation}
		\bs{F}_t= \bs{I}+ \bs{\varepsilon}_{t}=
\bs{I}+\sum_{i=1}^m {\bs R}_g\cdot \tilde{\bs{\varepsilon}}_{ti} \cdot {\bs R}_g^T  c_i,
		\label{tr-str}
	\end{equation}
 where  $\bs {\varepsilon}_{t}$ is the  transformation strain, $\tilde{\bs{\varepsilon}}_{ti}$ is the transformation strain of the $i^{th}$ martensitic variant
 in the local crystallographic basis of the grain,
 and the $c_i$ is the volume fraction of the $i^{th}$ variant in terms of volumes in the reference configuration.
	
	The  total, elastic, and transformational  Lagrangian strains  are defined as
		\begin{equation}
		\bs{E}=\frac{1}{2}(\bs{F}^T{\bs{\cdot}}\bs{F}-\bs{I});\quad
		\bs{E}_e=\frac{1}{2}(\bs{F}_e^T{\bs{\cdot}}\bs{F}_e-\bs{I});\quad
		\bs{E}_t=\frac{1}{2}(\bs{F}_t^T{\bs{\cdot}}\bs{F}_t-\bs{I}). \qquad
		\label{Esall}
	\end{equation}
Utilization of multiplicative decomposition \cref{Fdecom}	results in the following relationship:
\begin{equation}
\bs{E}_e = \bs{F}_t^{-1}{\bs{\cdot}}		(\bs{E}-\bs{E}_t){\bs{\cdot}} \bs{F}_t^{-1}.
		\label{Erel}
	\end{equation}
	
We will also need the ratios of elemental volumes $dV$ and mass densities $\rho$ in the different configurations, which are described by the Jacobian determinants:
	\begin{equation}
	    J=\frac{d V}{dV_0}=\frac{\rho_0}{\rho}=\det{\bs F}; \quad J_t=\frac{d V_t}{dV_0}=\frac{\rho_0}{\rho_t}=\det{\bs F_t}; \quad J_e=\frac{d V}{dV_t}=\frac{\rho_t}{\rho}=\det{\bs F_e}.
	\end{equation}

	\subsection{Helmholtz Free energy}
	In defining the Helmholtz free energy $ \psi $ per unit undeformed volume in $\Omega_0$ of the mixture of austenite and $ m $ martensitic variants, contributions from the elastic $ \psi^{e} $, thermal $ \psi_i^{\theta} $, and interaction $ \psi^{in} $ energy are  given by
	\begin{equation}
		\psi(\bs{F}_e,c_i,\theta)= J_t \psi^{e}(\bs{F}_e,c_i)+\psi^{\theta}(\theta,c_i)+\psi^{in}(c_i).
	\end{equation}
	Here, $ \psi^{e} $ is defined in $\Omega_t$, and the Jacobian $J_t$ maps it into  $\Omega_0$; $ \psi_i^{\theta} $ includes the thermal driving force for the PT and depends on the temperature $ \theta $; $ \psi^{in} =\mathcal{A} c c_0 \geq 0$ includes the interactions between austenite and martensite, the energy of internal stresses, as well as the
austenite-martensite phase interface energy; interaction  between martensitic variants is neglected to avoid formation of
variant-variant interfaces. Positive $\mathcal{A}$ results in a negative tangent modulus of the equilibrium stress-strain curve during
PT, which results in local mechanical instability and the formation of the localized transformation bands/regions of the product phase. In such a way, a discrete martensitic structure is reproduced, similar to the nanoscale PFA.
	
	
	The elastic energy is expressed as
	\begin{equation}\label{elastic_energy}
		\psi^e=\frac{1}{2}\fg {E}_e \fg :\tilde{\fg C} \fg : \fg {E}_e =\frac{1}{2}\tilde{C}^{ijkl} {E}_e^{ij} {E}_e^{kl};
\qquad \tilde{C}^{ijkl}=\sum_{p=0}^{m} \tilde{C}^{ijkl}_pc_p,
	\end{equation}

	\noindent where the components of the forth-rank elastic moduli tensor $\tilde{C}^{ijkl}_p$ in the global coordinate system
are defined in terms of components in the local for each grain crystallographic system ${C}^{rqnm}_p$	 and grain rotations $R_g^{ml}$ as
	\begin{equation}
\tilde{C}^{ijkl}_p=R_g^{lm}R_g^{kn}R_g^{jq}R_g^{ir}{C}^{rqnm}_p.
		\label{C_rotation}
	\end{equation}
	
	Since the thermal energy of all martensitic variants is the same, $\psi_i^{\theta}=\psi_j^{\theta}=\psi_M^{\theta}$,
the thermal energy of the mixture is
	\begin{equation}\label{thermal_energy}
		\psi^\theta=\sum_{i=0}^{m}c_i\psi_i^{\theta}(\theta)=c_0\psi_A^\theta(\theta)+c\psi_M^\theta(\theta);
\qquad c = \sum_{i=1}^{m}c_i; \qquad c_0 = 1-c,
	\end{equation}
	where $ \psi_A^\theta $ is  the thermal energy of austenite, $c$ and $c_0$  are the volume fractions of the martensite and austenite.
	
	\subsection{Dissipation inequality}
	\par The  Plank's inequality for isothermal processes is  
	\begin{equation}\label{general_dissipation}
		D=\bs{P}^T \bs{:}\dot{\bs{F}}-\dot{\psi}\geq 0,
	\end{equation}
	where $ D $ is the dissipation rate per unit undeformed volume; $ \bs{P} $ is the first Piola-Kirchhoff stress.
After traditional thermodynamic manipulations, $D$ can be expressed as the product of the  thermodynamic driving forces for $ A\rightarrow M_i  $, $X_{i0}$, and  $ M_j\rightarrow M_i $,  $X_{i0}$, transformations and conjugate
rates:
	\begin{align}\label{dissipation}
		\nonumber
		&D=\sum_{i=1}^{m}X_{i0}\dot{c}_{i0}+\sum_{j=1}^{m-1}\sum_{i=j+1}^{m}X_{ij}\dot{c}_{ij} \geq 0;\\\nonumber
		&X_{i0}=W_{i0}-\frac{J_t}{2}\bs{E}_e\bs{:}(\bar{\bs{C}}_i-\bar{\bs{C}}_0)\bs{:}\bs{E}_e-
		\frac{J_t}{2}\left(\bs{E}_e\bs{:}\bar{\bs{C}}(c)\bs{:}\bs{E}_e \right) {\fg F}_{t}^{-1} \bs{:} \bs{\varepsilon}_{ti}
		-\Delta \psi^{\theta}-\mathcal{A}(1-2c); 
		\\\nonumber
		&X_{ij}=W_{ij}-\frac{J_t}{2}\bs{E}_e\bs{:}(\bar{\bs{C}}_i-\bar{\bs{C}}_j)\bs{:}\bs{E}_e
		-\frac{J_t}{2}\left( \bs{E}_e\bs{:}\bar{\bs{C}}(c)\bs{:}\bs{E}_e \right) {\fg F}_{t}^{-1} \bs{:} (\bs{\varepsilon}_{ti}-\bs{\varepsilon}_{tj})
		;
		\\
		&W_{i0} =\bs{P}^T\bs{\cdot}\bs{F}_e\bs{:}\bs{\varepsilon}_{ti}=
		J{\fg F}^T_e {\fg \cdot} \fsg
		\bs{\cdot} {\fg F}_{e}^{t-1}  \cdot {\fg F}_{t}^{-1} \bs{:}\bs{\varepsilon}_{ti};
		\quad
		W_{ij}= \bs{P}^T\bs{\cdot}\bs{F}_e\bs{:}(\bs{\varepsilon}_{ti}-\bs{\varepsilon}_{tj})=
		J {\fg F}^T_e {\fg \cdot} \fsg
		\bs{\cdot} {\fg F}_{e}^{t-1}  \cdot {\fg F}_{t}^{-1}  \bs{:}(\bs{\varepsilon}_{ti}-\bs{\varepsilon}_{tj}).
	\end{align}
	Here, $\dot{c}_{i0}$ and $\dot{c}_{ij}$ are the rate of change of volume fraction of variant $i$ due to transformation to the austenite and variant $j$, respectively; $W_{i0}$ is the transformation work for  austenite to martensite PT,  $W_{ij}$ is the transformation work for variant $j$ to variant $i$ transformation,   $\Delta \psi^{\theta}$ is the jump in thermal energy during  transformation, and $\fsg=J^{-1}  \bs{P} \bs{\cdot}\bs{F}^T $ is the Cauchy (true) stress tensor. 

	\subsection{Kinetic equations}
	The kinetic equations are formulated as follows for the $ A\leftrightarrow M_i$ PTs
	\begin{equation}\label{kinetic_i0}
\begin{cases}
\begin{split}
\dot{c}_{i0}=\lambda_{i0}(X_{i0}-k_{i-0}) \quad \text{if}\;
&\{X_{i0}-k_{i-0}(c_i, \sg_i)>0 \;\&\; c_i<1 \;\&\; c_0>0\} \; \qquad A \rightarrow M_i\\
\dot{c}_{i0}=\lambda_{i0}(X_{i0}+k_{i-0}) \quad
\text{if}\; &\{X_{i0}+k_{i-0}(c_i, \sg_i)<0 \;\&\; c_i>0 \;\&\; c_0<1\}\qquad \; M_i \rightarrow A
\end{split}\\
\dot{c}_{i0}=0 \qquad \qquad \qquad \; \text{otherwise;} \qquad i=1,2,...,m,
\end{cases}
\end{equation}
and for $ Mj\leftrightarrow M_i $ PTs
\begin{equation}\label{kinetic_ij}
\begin{cases}
\begin{split}
\dot{c}_{ij}=\lambda_{ij}X_{ij} \quad \text{if}\;
&\{X_{ij}>0 \;\&\; c_i<1 \;\&\; c_j>0\}\qquad j \rightarrow i\\
\text{or}\; &\{X_{ij}<0 \;\&\; c_i>0 \;\&\; c_j<1\} \qquad i \rightarrow j
\end{split}\\
\dot{c}_{ij}=0 \qquad \qquad \qquad \qquad \qquad \quad \text{otherwise;} \qquad i,j = 1,2,...,m,
\end{cases}
\end{equation}
where  $k_{i-0}$ is the  athermal threshold and  $ \lambda_{i0} $ and $ \lambda_{ij} $ are the kinetic coefficients.
We neglect the athermal threshold for $ Mj\leftrightarrow M_i $ PTs.
The non-strict inequalities for the volume fraction of phases in   Eqs. (\ref{kinetic_i0})-(\ref{kinetic_ij})
imply  that the PT from any phase does not occur if the parent phase does not exist or  if the product phase is complete.	
	
	\subsection{Macroscopic parameters}\label{Macroscopic}
Macroscopic Cauchy stress and the first Piola-Kirchhoff stress, the deformation gradient, transformation strain, and volume fraction of   Si II and each martensitic variant, averaged over the sample, are defined as \cite{Hill1984OnStrain,Levitas1996SomeSurfaces,Petryk1998MacroscopicTransformation}
 \bey
\bar{\fsg}=\frac{1}{V}\int_V \fsg dV; \qquad \bar{\fg P}=\frac{1}{V_0}\int_{V_0} {\fg P} dV_0;
  \label{av-stress}
 \eey
 \bey
\bar{\fg F}=\frac{1}{V_0}\int_{V_0} \fg F dV_0; \qquad
\bar{\fg F}_t\simeq\frac{1}{V_0}\int_{V_0} \fg F_t dV_0;
\qquad  \bar{\fvep}_t \simeq \frac{1}{V_0}\int_{V_0} {\fvep}_t dV_0;
  \label{av-Fp}
 \eey
 \bey
\bar{c}=\frac{1}{V_0}\int_{V_0} c dV_0;
\qquad  \bar{c}_i=\frac{1}{V_0}\int_{V_0} c_i dV_0.
  \label{av-c}
 \eey
	For the Cauchy stress and the first Piola-Kirchhoff stress, the deformation gradient, and volume fractions $c$ and $c_i$, averaging equations are strict; for the transformation deformation gradient and strain, they are approximate because, in the unloaded stress-free state, they generally
are not compatible due to residual elastic strain. We use Eq. (\ref{av-Fp}) because the exact equation is quite bulky and is for $\dot{\fg F}_t$ instead of  ${\fg F}_t$. Eq. (\ref{av-c}) for $c_i$, while formally correct, is misleading (\Cref{Strain-controlled}).  Other macroscopic parameters are defined via $\bar{\fg F}$
and $\bar{\fg P}$ by equations similar to the corresponding local equations:
 \bey
\bar{\fsg}=(\det\bar{\fg F})^{-1}  \bar{\bs{P}} \bs{\cdot}\bar{\bs{F}}^T;
\qquad
\bar{\bs{E}}=\frac{1}{2}(\bar{\bs{F}}^T{\bs{\cdot}}\bar{\bs{F}}-\bs{I}).
  \label{av-sg-E}
 \eey
Eq. (\ref{av-sg-E}) for the Cauchy stress is used instead of     Eq. (\ref{av-stress}) because integration over the fixed
parallelepiped with regular cubic mesh is much simpler and faster than the integration over the deformed volume.

	\section{Model parameters}\label{Model parameters} 
The material parameters required for the implementation of the model, the same as in \cite{babaei2020finite}, are presented in \cref{model} are listed in \cref{tab1}. The transformation strain tensors  are taken from the MD simulations in {\cite{Levitas2017LatticeCriterion,Levitas2017Triaxial-Stress-InducedPhases}}, and in the local crystallographic axes for all three variants are given by
\begin{align}
\scalebox{0.9}{$
	\tilde{\fvep}_{t1}=
	\begin{pmatrix}
		0.1753 & 0 & 0 \\
		0 & 0.1753 & 0 \\
		0 & 0 & -0.4470
	\end{pmatrix};
	\quad
	\tilde{\fvep}_{t2}=
	\begin{pmatrix}
		0.1753 & 0 & 0 \\
		0 & -0.4470 & 0 \\
		0 & 0 & 0.1753
	\end{pmatrix};
	\quad
	\tilde{\fvep}_{t3}=
	\begin{pmatrix}
	-0.4470	& 0 & 0 \\
		0 & 0.1753 & 0 \\
		0 & 0 &0.1753
	\end{pmatrix}.
$}
\end{align}
{The elastic constants for both phases are collected  from \cite{Schall-etal-2008,Chen-etal-NPJ-2020}}. The constants $C_0^{ij}$ in \cref{tab1} denote the independent elastic constants of the austenite, and ${C}_1^{ij}$ denote those of the first variant of the martensite, both in the local crystallographic axes. The components of the tensor of elastic moduli $C^{ijkl}$ can be calculated using
\begin{eqnarray}
	\nonumber
	C^{ijkl}=\sum_{n=1}^{3}[\lambda^n\delta^{in}\delta^{jn}\delta^{kn}\delta^{ln}+\mu^n(\delta^{in}\delta^{jn}
	\delta^{kl}+\delta^{ij}\delta^{kn}\delta^{ln})\\
	+\nu^n(\delta^{in}\delta^{jk}\delta^{ln}+\delta^{jn}\delta^{ik}\delta^{ln}+\delta^{in}\delta^{jl}\delta^{kn}
	+\delta^{jn}\delta^{il}\delta^{kn})],
\end{eqnarray}
where $\lambda^n$, $\mu^n$ and $\nu^n$ for a cubic and tetragonal crystal lattice are given by \cref{cubic,tetragonal}, respectively:
\begin{align}{\label{cubic}}
	\nonumber
	&\lambda^1=\lambda^2=\lambda^3=C^{11}-C^{12}-2C^{44},\\\nonumber
	&2\mu^1=2\mu^2=2\mu^3=C^{12},\\
	&2\nu^1=2\nu^2=2\nu^3=C^{44}.
\end{align}
\begin{align}\label{tetragonal}
	\nonumber
	&\lambda^1=\lambda^2=C^{11}-(C^{12}+2C^{66}),\\\nonumber
	&\lambda^3=C^{33}+C^{12}+2C^{66}-2(C^{13}+2C^{44}),\\\nonumber
	&2\mu^1=2\mu^2=C^{12}, \qquad \;\;2\mu^3=2C^{13}-C^{12},\\
	&2\nu^1=2\nu^2=C^{66},\qquad \;\; 2\nu^3=2C^{44}-C^{66}.
\end{align}

Under hydrostatic conditions, the phase equilibrium pressure $p^{eq}_0= 10.5 GPa$ {\cite{Voronin-etal-2003}}  at $J_t=0.764$.   The jump in the thermal energy  $\Delta\psi^\theta=-p^{eq}_0({J}_t-1)=2.47 GPa$, where   elastic strain and change in elastic moduli are neglected.
Localization of strain is required to reproduce discrete microstructure. As noted in \cite{babaei2020finite}, to obtain strain localization, the strain rate should  be commensurate with the rate of transformation. The kinetic coefficient $\lambda$ and interaction parameter $A$ are selected to ensure that this condition is satisfied.

It is found with the first principle and molecular dynamics simulations \cite{zarkevich2018lattice,Levitas2017LatticeCriterion,Levitas2017Triaxial-Stress-InducedPhases} that the criteria for
 for Si I$\leftrightarrow$Si II PTs are linearly dependent on the  Cauchy stress components normal to the cubic faces.
We assume the same for  the microscale experiments, where the role of defects is effectively included. For cubic to tetragonal PTs, the PT criteria are given by~\citep{Babaei-Levitas-ActaMat-2019,babaei2020finite}
\begin{align}\label{experimental_PT}
	\nonumber
	&A\rightarrow M_i:\qquad a^d(\sigma_{1}+\sigma_{2})+b^d \sigma_{3}>c^d;\\ &M_i\rightarrow A:\qquad a^r(\sigma_{1}+\sigma_{2})+b^r \sigma_{3}<c^r,
\end{align}
where $ a^d,b^d,c^d,a^r,b^r $ and $ c^r $ are constants determined empirically.  To make the thermodynamic PT conditions consistent with experimental conditions, the athermal threshold $k_{i-0}$ is considered to be stress and volume fraction dependent, as described in detail in  ~\citep{babaei2020finite}, and calculated based on the relations
\begin{equation}\label{k_i0}
	k_{i-0}=J[a_1(c_i) (\sigma_{1}+\sigma_{2})+a_3(c_i)\sigma_{3}]; \qquad
	a_k(c_i)=d_k+(r_k-d_k)c_i; \;\;k=1\; \rm{ and} \; 3,
\end{equation}
where $d_k$ and $r_k$ are the fitting parameters, given in \cref{tab1}.
They are determined by substituting expressions for $	k_{i-0}$ from \cref{k_i0} in the transformation criteria in
 \cref{kinetic_i0} with the driving forces from
\cref{dissipation}  and some empirical data.

\begin{table}[h]
	\centering
	\caption{
		Material parameters including kinetic coefficient $\lambda \, \rm{(Pa \cdot s)}^{-1}$, dimensionless constants in the
		expression for effective thresholds, as well as interaction coefficient $A$, jump in the thermal energy
		$ \Delta \psi^{\theta} $,  and elastic constants, all in GPa.
	}
	\begin{tabular}{c c c c c c c c c }
		\hline
		$\lambda$   & $A$ & $\Delta\psi^\theta $  & $d_1$&$d_3$&$r_1$&$r_3$
		\\
		\hline
		0.02 & 2 & 2.47 & 0.082 &0.111&-0.90&0.338 \\
		\hline
		$ C_0^{11}$  & $C_0^{44}$ & $C_0^{12}$ & $C_1^{11}$ & $C_1^{33}$ & $C_1^{44}$ & $C_1^{66}$ & $C_1^{12}$ & $C_1^{13}$
		\\
		\hline
		167.5 & 80.1 & 65.0 & 174.76 & 136.68 & 60.24 & 42.22 & 102.0 & 68.0\\
		\hline
	\end{tabular}
	\label{tab1}
\end{table}

\begin{figure}
	\centering
	\resizebox{150mm}{!}{\includegraphics[trim={5mm 100mm 100mm 45mm},clip]{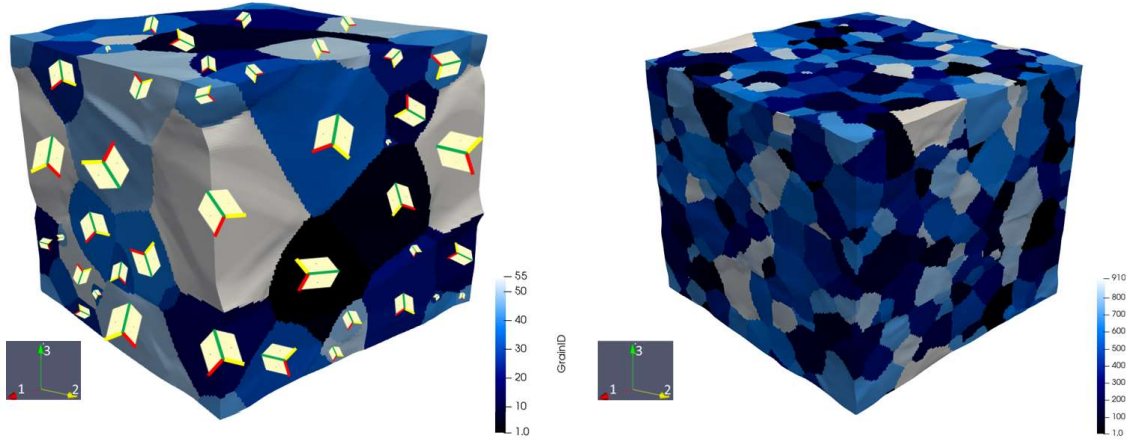}}
	\caption{Grain distribution generated from DREAM.3D for (a) 55 grains showing the local orientations of individual grains (red=1, yellow=2, green=3) and (b) 910 grains. Each grain has a different orientation that is randomly assigned to make sure the sample, on the whole, is texture-free. Files with complete information about orientation of each grain for both samples are given in supplementary material.}
	\label{grainID}
\end{figure}

\section{Study of multivariant microstructure evolution}\label{results}
A    finite element implementation of the scale-free model is developed in the open-source FEM code deal.II~\citep{Bangerth2007Deal.IIALibrary} using 8-noded 3D cubic linear elements with first-order interpolation and full integration. For such elements, calculations of volume averaged of any parameter $a$ in the undeformed configuration is the sum of values $a$ in each quadrature point divided by the number of quadrature points.
Microstructure evolution is studied for polycrystalline samples containing two different numbers of grains,  55 and 910, as shown in \cref{grainID}.
The total number of the finite elements and integration points were $\simeq$2.1 million and $\simeq$16.7 million, respectively, for both cases.
If we consider position vector $\fg r$ and volume fractions $c_1$, $c_2$, and $c_3$ as the primary independent variables, then the total number of degrees of freedom is six times the number of the integration points.

A 3D unit cube sample is constructed in DREAM.3D~\citep{groeber2014dream} with the required number of grains. Each grain in the sample is assigned an orientation randomly so that the overall sample remains texture free. Two types of compression are applied, strain-controlled and stress-controlled.
For both cases, periodic boundary conditions are applied in directions 1 and 2 with averaged zero averaged
strains in these directions. Application of the periodic boundary conditions significantly simplified elimination of divergence of the computational procedure in comparison with other conditions. Such conditions are realized when Si is clamped in the 1-2 plane; a thin Si layer is attached to the rigid substrate, or within a shock wave.
For  the strain-controlled case, periodic boundary conditions are applied in the direction 3 as well, and a  sample is subjected to a uniaxial compressive averaged strain in direction 3.  For the stress-controlled loading, on the external planes orthogonal to axis 3, the sample is subjected to homogeneous  compressive normal Cauchy stress  in the deformed configuration with zero shear stresses, i.e., like under the action of the liquid. 
Stress-controlled
loading ends with constant "pressure in liquid " of 11 GPa.
In total, four simulations are run with two different numbers of grains and two different loading conditions.


\subsection{Strain-controlled loading}\label{Strain-controlled}
\begin{figure}
	\centering
	\resizebox{150mm}{!}{\includegraphics[]{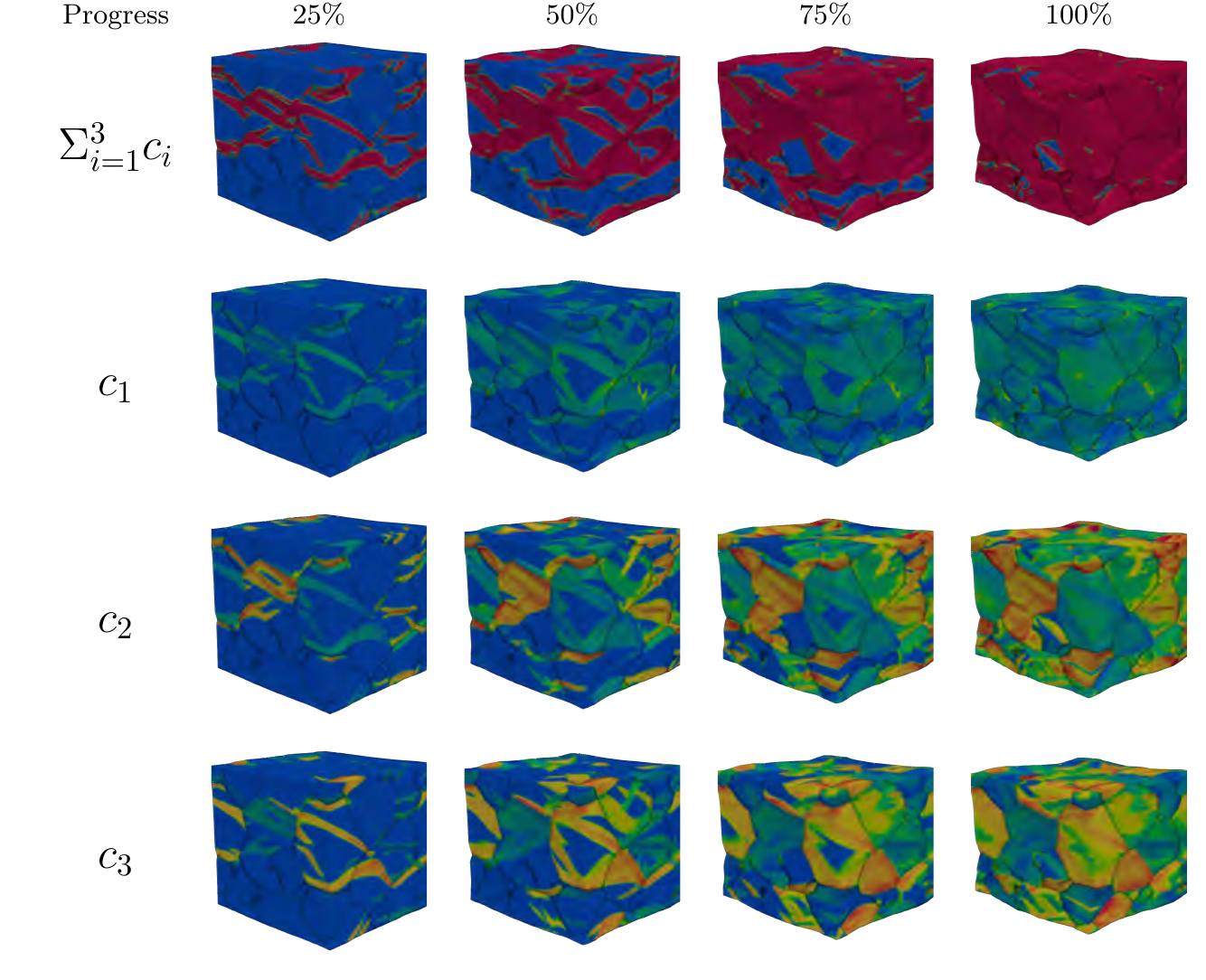}}
	\begin{minipage}{\linewidth}
	\centering
	\hspace{0.05\linewidth}
	\begin{minipage}{0.2\linewidth}
		\begin{figure}[H]
			\resizebox{20mm}{!}{\includegraphics{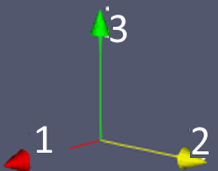}}
		\end{figure}
	\end{minipage}
	\hspace{0.05\linewidth}
	\begin{minipage}{0.5\linewidth}
		\begin{figure}[H]
			\resizebox{60mm}{!}{\includegraphics{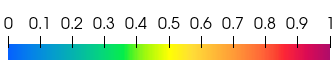}}
		\end{figure}
	\end{minipage}
\end{minipage}
	\caption{Evolution of volume fractions of phases for strain-controlled loading of a sample with 55 grains. The figure shows snapshots at different stages of completion of simulation  (25\%, 50\%, 75\%, and 100\%) in different columns  for Si II, $c$, and each martensitic variant, $c_i$.}
	\label{55_cevol}
\end{figure}

\begin{figure}
	\centering
	\resizebox{150mm}{!}{\includegraphics[]{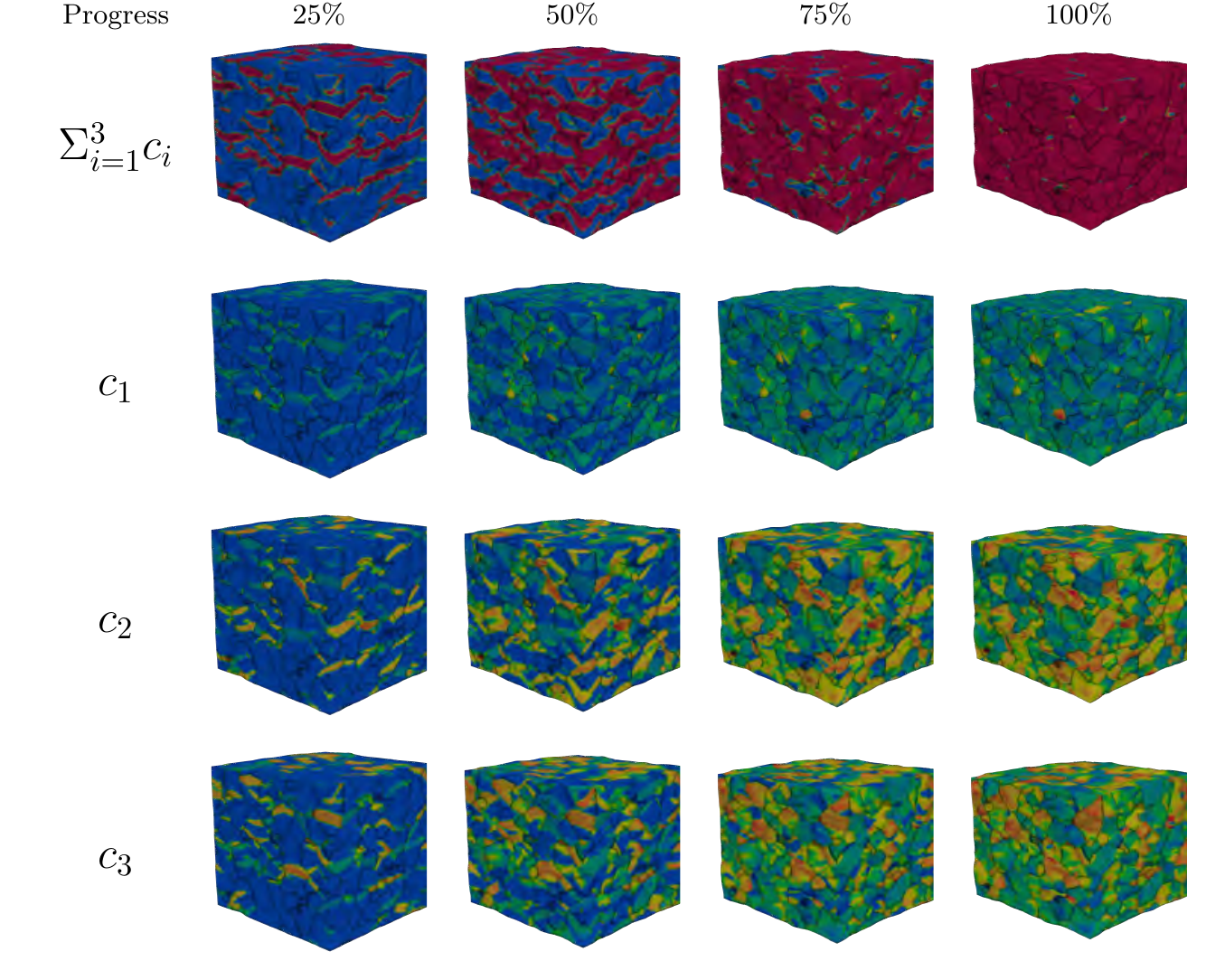}}
	\begin{minipage}{\linewidth}
	\centering
	\hspace{0.05\linewidth}
	\begin{minipage}{0.2\linewidth}
		\begin{figure}[H]
			\resizebox{20mm}{!}{\includegraphics{Images/axes.png}}
		\end{figure}
	\end{minipage}
	\hspace{0.05\linewidth}
	\begin{minipage}{0.5\linewidth}
		\begin{figure}[H]
			\resizebox{60mm}{!}{\includegraphics{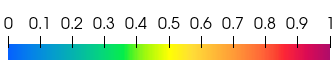}}
		\end{figure}
	\end{minipage}
\end{minipage}
	\caption{Evolution of volume fractions for strain-controlled loading of a sample with 910 grains. }
	\label{910_cevol}
\end{figure}

\begin{figure}[H]
			\centering
			\resizebox{70mm}{!}{\includegraphics{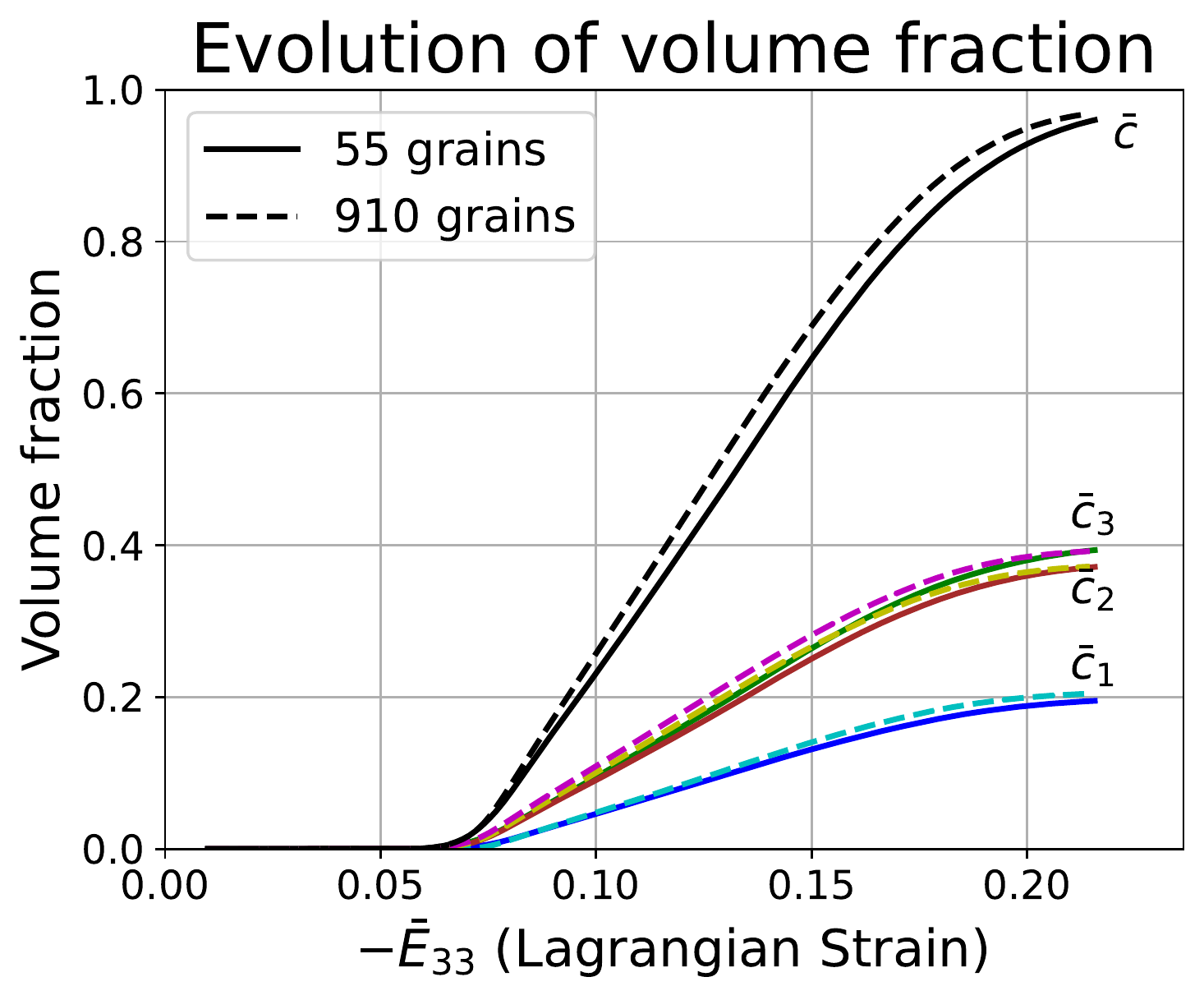}}
			\caption{Volume fraction of each martensitic variant ($\bar{c}_i$) and the total martensite ($\bar{c}$) averaged over the sample based on Eq. (\ref{av-c})
vs. strain  for strain-controlled loading.
The lack of physical sense for $\bar{c}_i$ is described in the text.}
			\label{strain_vf}
		\end{figure}

For  strain-controlled loading, the initial strain rate used, in the elastic regime, is $1\times 10^{-2} s^{-1}$ and reduced to $5\times 10^{-5} s^{-1}$ once the material starts to transform. While we can choose any strain rate, such low strain rates are chosen because they are not achievable in atomistic simulations. 
 \Cref{55_cevol,910_cevol} show the evolution of martensite at three intermediate stages (25\%, 50\%, and 75\% of the simulation) and the final stage of the simulation. The first row shows the total volume fraction of martensite $c$ followed by those for individual martensitic variants. The black lines in \cref{55_cevol,910_cevol} outline the grain boundaries for all the grains.  The videos demonstrating the evolution of the volume fractions are provided in the supplementary material.
 Volume fractions of each martensitic variant $\bar{c}_i$ and martensite $\bar{c}$ averaged over the sample based on Eq. \ref{av-c}
vs. strain are shown in Fig.	\ref{strain_vf}.
Nucleation starts mostly at the grain boundaries and triple junctions, which represent stress concentrators.
Nucleation starts with the dominant variant, which produces maximum transformation work;
two other variants appear in the same regions to accommodate the transformation strain and reduce internal stresses.
A similar proportion between variants remains during further growth.
In grains with different orientations, growth occurs either along the grain boundaries or inside the grain, which is arrested
either at grain boundaries or other martensitic units. In each unit full transformation to Si II occurs quickly, thus
forming discrete Si II microstructure, as desired. With increasing strain, both Si II broadening of existing Si II regions and the appearance of new nuclei occur. At the end of the loading, PT is completed almost everywhere, with small residual Si I pockets. They are caused by internal stresses due to large transformation strains. For both  numbers of grains, $\bar{c}$ saturates at 96\%.

   It can be clearly seen from 	\cref{55_cevol,910_cevol,strain_vf}  that the second and the third variants are dominating over the first one, which looks contradictory.
  To better understand the reasons for these results, let, for simplicity and illustration, consider 5 grains:
  grain 1  oriented with $[100]$ direction along  axis 3 (variant 1); grains 2 and 3, rotated by $\pm 90^o$ about axis 1 (variant 1), and
   grains 4 and 5, rotated by $\pm 90^o$ about axis 2 (variant 3). Due to symmetry, this is still the same single crystal.
      If we treat it like a single crystal and transform it homogeneously till completion, we obtain $ c_1=1$, $c_2=c_3=0$, and $\fvep_t=\fvep_{t1}$.
       However, if we treat it as a polycrystal, in grain 1, we will have the same $ c_1=1$, $c_2=c_3=0$ and $\fvep_t=\fvep_{t1}$.
In the local coordinate system of grains 2 and 3,
      this transformation strain  looks like $\fvep_t=\fvep_{t2}$, i.e., $ c_2=1$, $c_1=c_3=0$.
      Similarly, in the local coordinate system of grains 4 and 5,
      the transformation strain  is $\fvep_{t3}$, i.e., $ c_3=1$, $c_1=c_2=0$.
  After averaging these volume fractions over the entire sample, we obtain     $ c_1=1/5$, $c_2=c_3=2/5$,  what we approximately observe in Fig.	\ref{strain_vf}.
   The main point is that ${c}_i$ for each grain does not have a lot of sense unless the orientation of the grain is shown.
   That is why we give complete information about orientation of each grain in supplementary material and show orientations  in \cref{grainID}a.
     The average over the sample $\bar{c}_i$ does not have any physical sense because the orientation of grains is not taken into account; they cannot be used in the macroscopic theories.  The average over the sample $\bar{c}$
     has clear physical sense.

\begin{figure}
	\centering
	\resizebox{150mm}{!}{\includegraphics[]{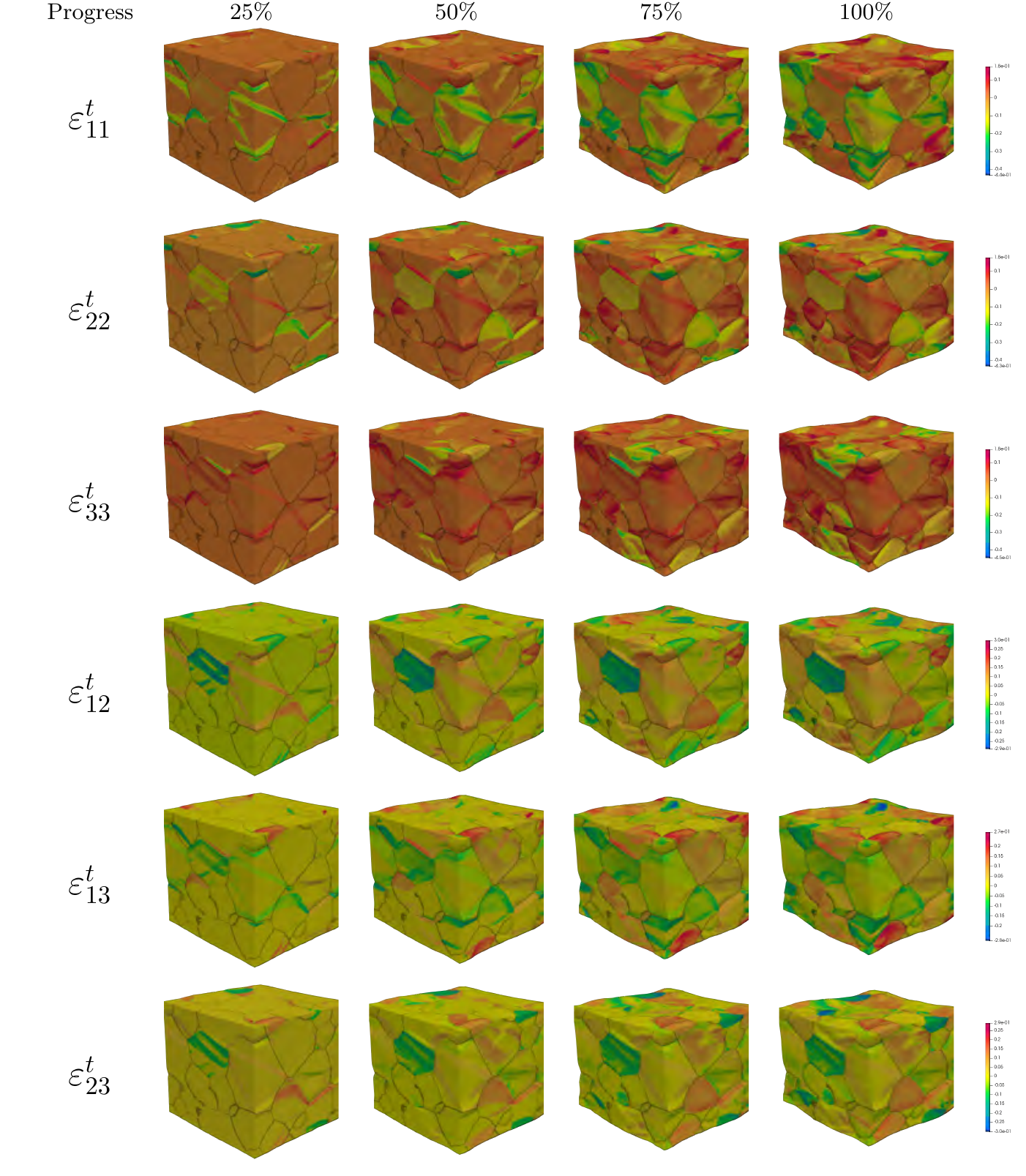}}
	\begin{minipage}{\linewidth}
	\hspace{0.1\linewidth}
	\begin{minipage}{0.1\linewidth}
		\begin{figure}[H]
			\resizebox{10mm}{!}{\includegraphics{Images/axes.png}}
		\end{figure}
	\end{minipage}
	\hspace{0.85\linewidth}
\end{minipage}
	\caption{Evolution of components of the transformation strain tensor $\fvep_t$  for strain-controlled loading of a sample with 55 grains. }
	\label{55_etevol}
\end{figure}

\begin{figure}
	\centering
	\resizebox{150mm}{!}{\includegraphics[]{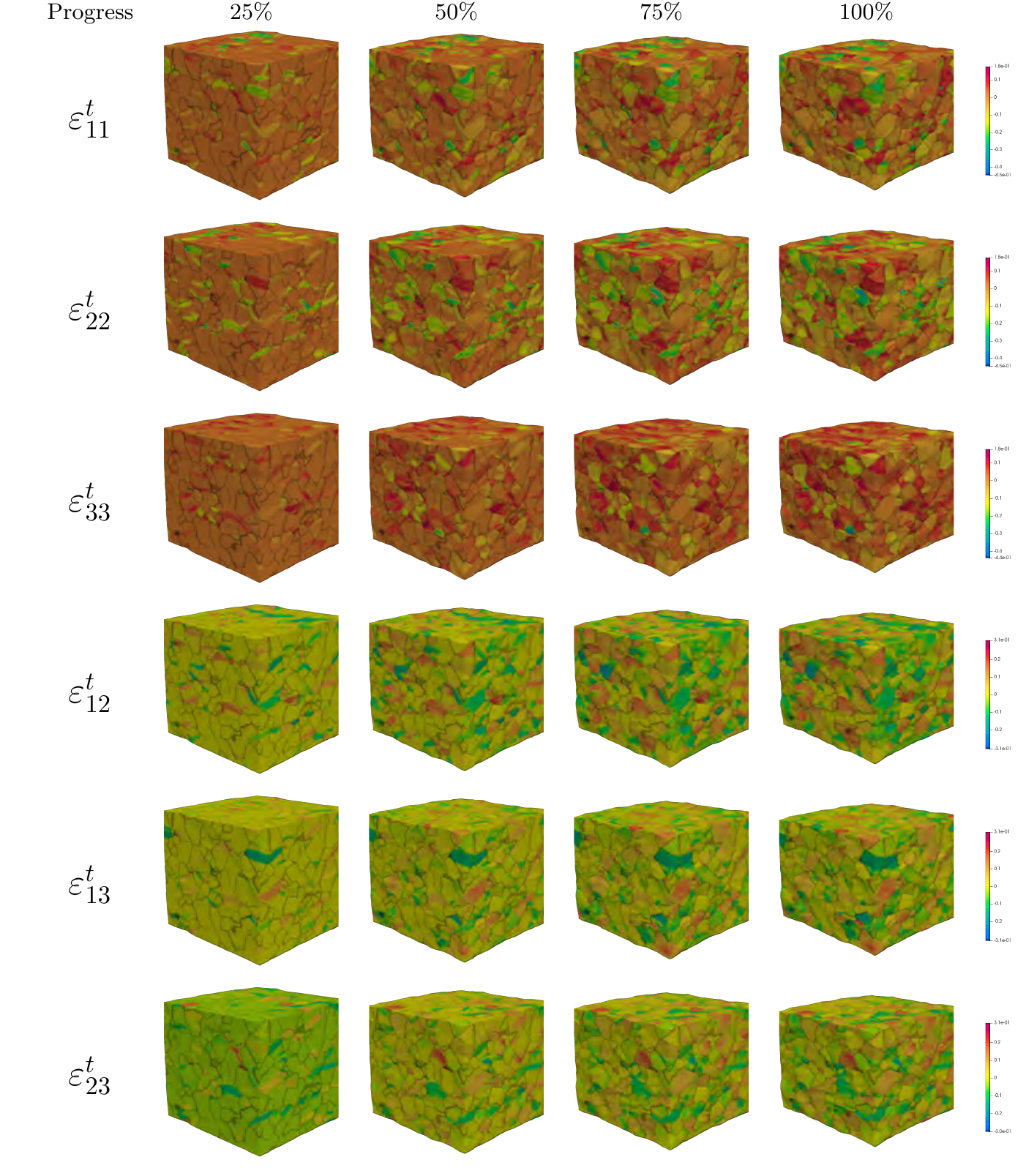}}
	\begin{minipage}{\linewidth}
	\hspace{0.1\linewidth}
	\begin{minipage}{0.1\linewidth}
		\begin{figure}[H]
			\resizebox{10mm}{!}{\includegraphics{Images/axes.png}}
		\end{figure}
	\end{minipage}
	\hspace{0.85\linewidth}
\end{minipage}
	\caption{Evolution of the components of the transformation strain tensor $\fvep_t$ for strain-controlled loading of a sample with 910 grains. }
	\label{910_etevol}
\end{figure}

     We suggest the following ways to present a multivariant structure in the polycrystal. For local presentation,
     one can show a triad of local crystallographic axes in each grain along with fields $c_i$ (\cref{grainID}a). One can present fields of  six components of
     $\fvep_{ti}c_i$   for each variant $M_i$ in the global coordinate system, which is 18 fields.
     For the above example with 5 grains, in each grain, we will have components of $\fvep_{t3}$ in the global coordinate system, which is consistent with the treatment of the aggregate as a single crystal.
     More compact is  to
     present six fields of  six components of the total transformation strain
     $\fvep_{t}=\sum_{i=1}^m \fvep_{ti}c_i$  in the global coordinate system (Figs. \ref{55_etevol}  and \ref{910_etevol}), which takes into account in more averaged sense both an orientation of grain and fields $c_i$.
       For the averaged description, one can use a plot of six components of $\bar{\fvep}_{t}$.

As it follows from Figs. \ref{55_etevol}  and \ref{910_etevol}, normal components of  $\fvep_{t}$ vary between extremes -0.447 and 0.1753 corresponding to the full local  transformation in single variant oriented along the global coordinate
axes, despite zero averaged total lateral strains. Shear components vary between extremes $~\pm 0.31$, also despite zero averaged.  Colors corresponding to zero value of each component of transformation strain is clear from colors of large Si I regions at 25\% transformation progress  corresponding $c=0$ in  Figs. \ref{55_cevol}  and \ref{910_cevol}. For both grain sizes one sees strong heterogeneity of transformation strains from grain to grain and within grains. Note that fields at the surface do not completely represent fields in the entire volume, that is why some results may look counterintuitive. For example, while averaged  $\vep_t^{33}$ is larger than $\vep_t^{11}$ and $\vep_t^{22}$,
this is not evident from the surface fields.

Despite the fact that for the smaller number of grains size of Si II units is larger,
difference in  $\bar{c} (\bar{E}_{33})$  (and even $\bar{c}_i (\bar{E}_{33})$)   for different number of grains is small (\cref{strain_vf}).
  Initiation of PT occurs at the same stress $\sg_{zz}=-10.33 $ GPa for both grain sizes, and stress-strain curves in  \cref{strain_con} also differ insignificantly. That means that the current model does not describe experimentally
  observed effect of the grain size on the Si I to Si II PT pressure or stress \citep{Sorb2022,zeng2020origin,xuan2020pressure,tolbert1996pressure}.
  The reason is that the current model does not include dislocations as local stress concentrators in bulk and at grain boundaries. This will be the next step in developing the current model. We can use the same approach for introducing discrete dislocations via the solution of the contact problem, as it was done in \cite{Levitas2018Scale-FreeMicrostructure,Esfahani-etal-2020} for small strains and 2D formulations. Of course, it is much more challenging to do this for 3D and large strains.

\begin{figure}
	\centering
	\resizebox{150mm}{!}{\includegraphics[]{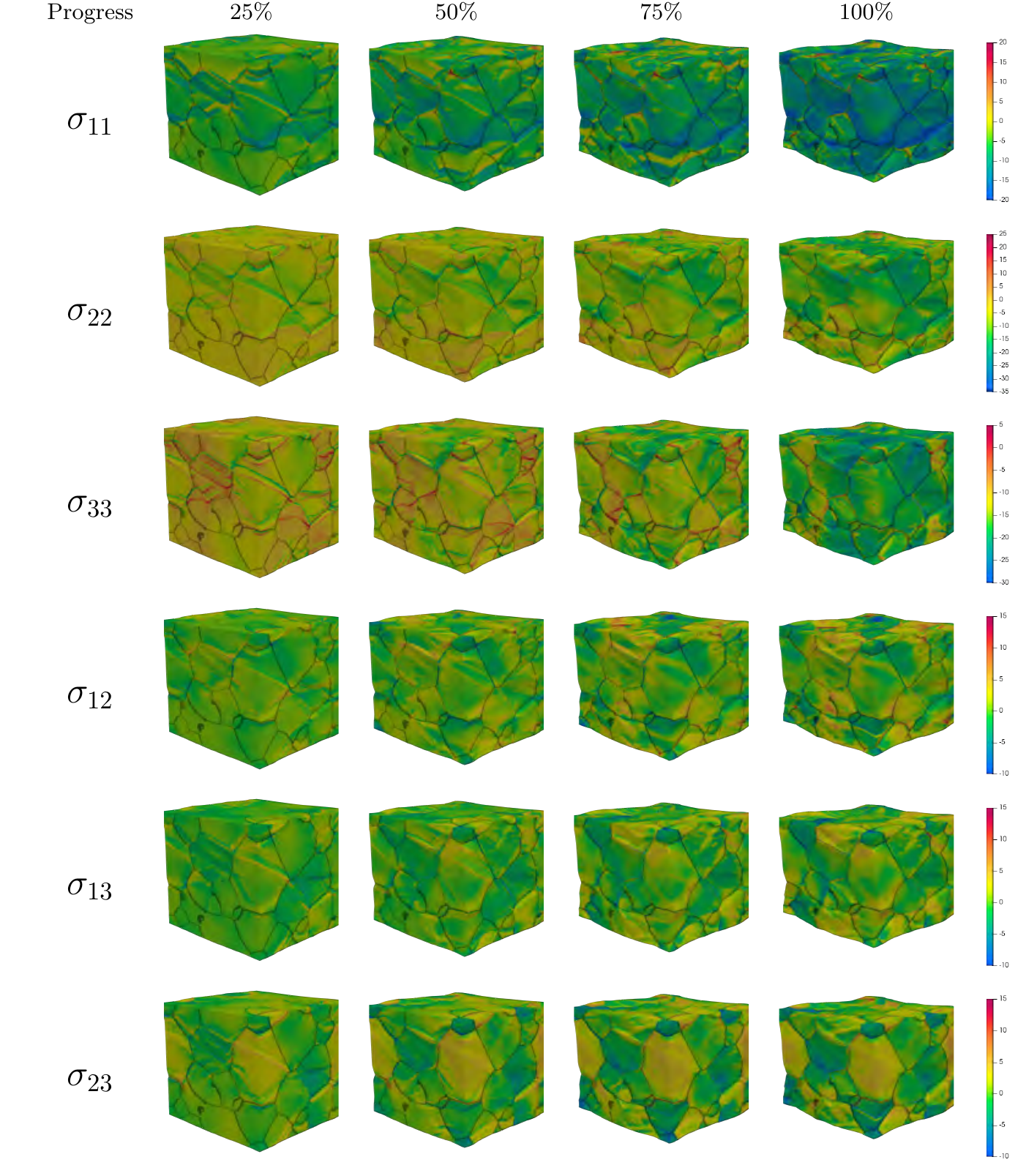}}
	\begin{minipage}{\linewidth}
	\hspace{0.1\linewidth}
	\begin{minipage}{0.1\linewidth}
		\begin{figure}[H]
			\resizebox{10mm}{!}{\includegraphics{Images/axes.png}}
		\end{figure}
	\end{minipage}
	\hspace{0.85\linewidth}
\end{minipage}
	\caption{Evolution of components of the Cauchy stress tensor for strain-controlled loading of a sample with 55 grains. }
	\label{55_sevol}
\end{figure}

\begin{figure}
	\centering
	\resizebox{150mm}{!}{\includegraphics[]{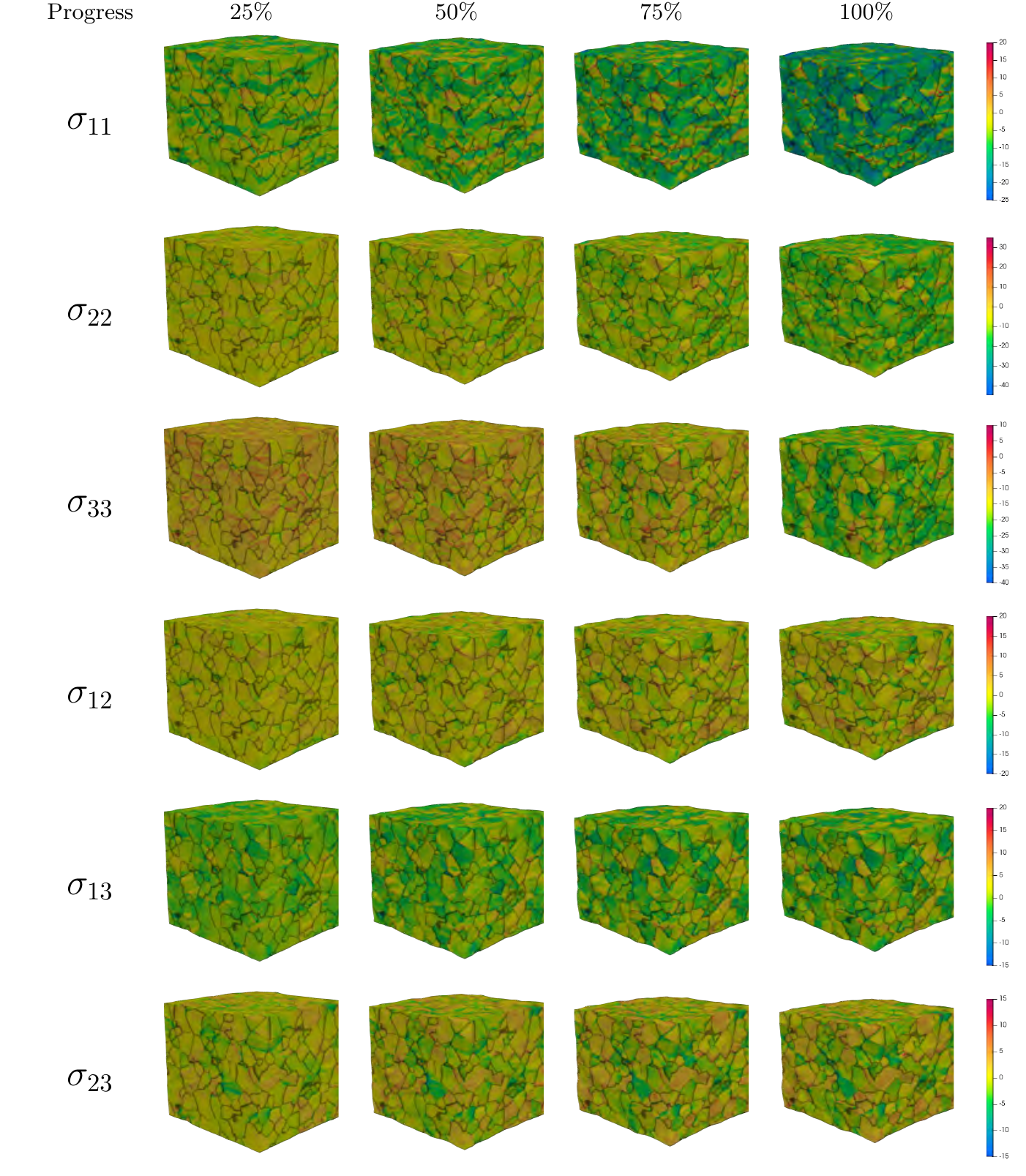}}
	\begin{minipage}{\linewidth}
	\hspace{0.1\linewidth}
	\begin{minipage}{0.1\linewidth}
		\begin{figure}[H]
			\resizebox{10mm}{!}{\includegraphics{Images/axes.png}}
		\end{figure}
	\end{minipage}
	\hspace{0.85\linewidth}
\end{minipage}
	\caption{Evolution of the components of the Cauchy stress tensor for strain-controlled loading of a sample with 910 grains. }
	\label{910_sevol}
\end{figure}

 \Cref{55_sevol,910_sevol} shows the evolution of all components of Cauchy stresses throughout the simulation. It can be noticed from \cref{55_sevol,910_sevol} that the grain boundaries and triple junctions have the highest  stresses of both signs, which cause nucleation of the dominating variant accompanied by two other variants
and growth, often along the grain boundaries. Next, a large stress concentration of both signs appears at the  phase interface, causing further nucleation in bulk (so-called autocatalytic effect).
The peak stresses are huge for both normal and even shear stresses. Thus, for small grains, $\sg_{22}$ varies from $- 45$ to $35$ GPa and $\sg_{33}$ varies from $- 40$ to $10$ GPa.
 Similar, shear stress $\sg_{12}$ varies from $- 20$ to $20$ GPa and $\sg_{23}$ varies from $- 15$ to $15$ GPa.
For large grains, the magnitude of extremes in stresses is smaller by $5$ to $10$ GPa.
Despite this difference for different grain sizes, the macroscopic PT initiation stress  $\sg_{zz}= -10.33$  GPa is the same, and the entire $\sg_{zz}-\vep_{zz}$ curve do not differ significantly. The initiation stress is determined by the transformation work, which depends on all components of the stress tensor. While stresses are different, the transformation work may be approximately the same for both grain sizes. This is similar to results in \cite{Babaei-Levitas-ActaMat-2019}, where a strong stress concentrator due to a dislocation in Si produced a relatively small contribution to the transformation work.

 The stress-strain plots for strain-controlled loading are given in \cref{strain_con}. After PT starts at $\bar{E}_{33}=-0.06$ and $\bar{\sg}_{33}=-10.33$ GPa for both small and large grains,  the stress-strain plots continue along the elastic curve  of the austenite with small nonlinearities. This behavior is because, initially, the transformation rate is very low, as can be noticed from \cref{strain_vf}.
 Si II nuclei are localized near stress concentrators without essential growth. Only at $\bar{E}_{33}=-0.07$ does intense growth start, which leads to a strong reduction in tangent modulus. Note that for a single crystal, the tangent modulus is getting negative at the onset of the PT, causing macroscopic instability
 \cite{babaei2020finite}. In contrast, for polycrystals, the local mechanical instabilities due to PT and negative local tangent modulus are stabilized at the macroscale by arresting/slowing the growth of Si II regions by the grain boundaries and generating the internal back stresses. This is reflected by the positive tangent moduli in the stress-strain plots in \cref{strain_con}. While intuitively, the more grain boundaries we have, the higher the tangent moduli should be, in fact, the response for 910 grains is slightly softer than that of 55 grains \cref{strain_con}. The reasons are:
 (a) more triple junctions and nucleation sites for smaller grains leading to Si II regions; (b) smaller misorientation between neighboring grains leading to easier transfer of PT growth from grain to grain, and (c) a larger number of  surrounding grains giving more chances to find proper orientation for new nucleation caused by internal stresses in the transforming grains. Although there is a noticeable difference in the response for both cases, the maximum difference is $<1$ GPa or $<10\%$. Such a small difference implies that it is unnecessary to treat a sample with such a large number of grains to estimate the macroscopic behavior of a polycrystal.	
  Note that the Lagrangian elastic strain at the end of simulation, at $\bar{E}_{33}=-0.22$,  is $~-0.1$, i.e., comparable to the transformation strain.

\begin{minipage}{\linewidth}
	\centering
	\begin{minipage}{0.45\linewidth}
		\begin{figure}[H]
			\centering
			\resizebox{70mm}{!}{\includegraphics{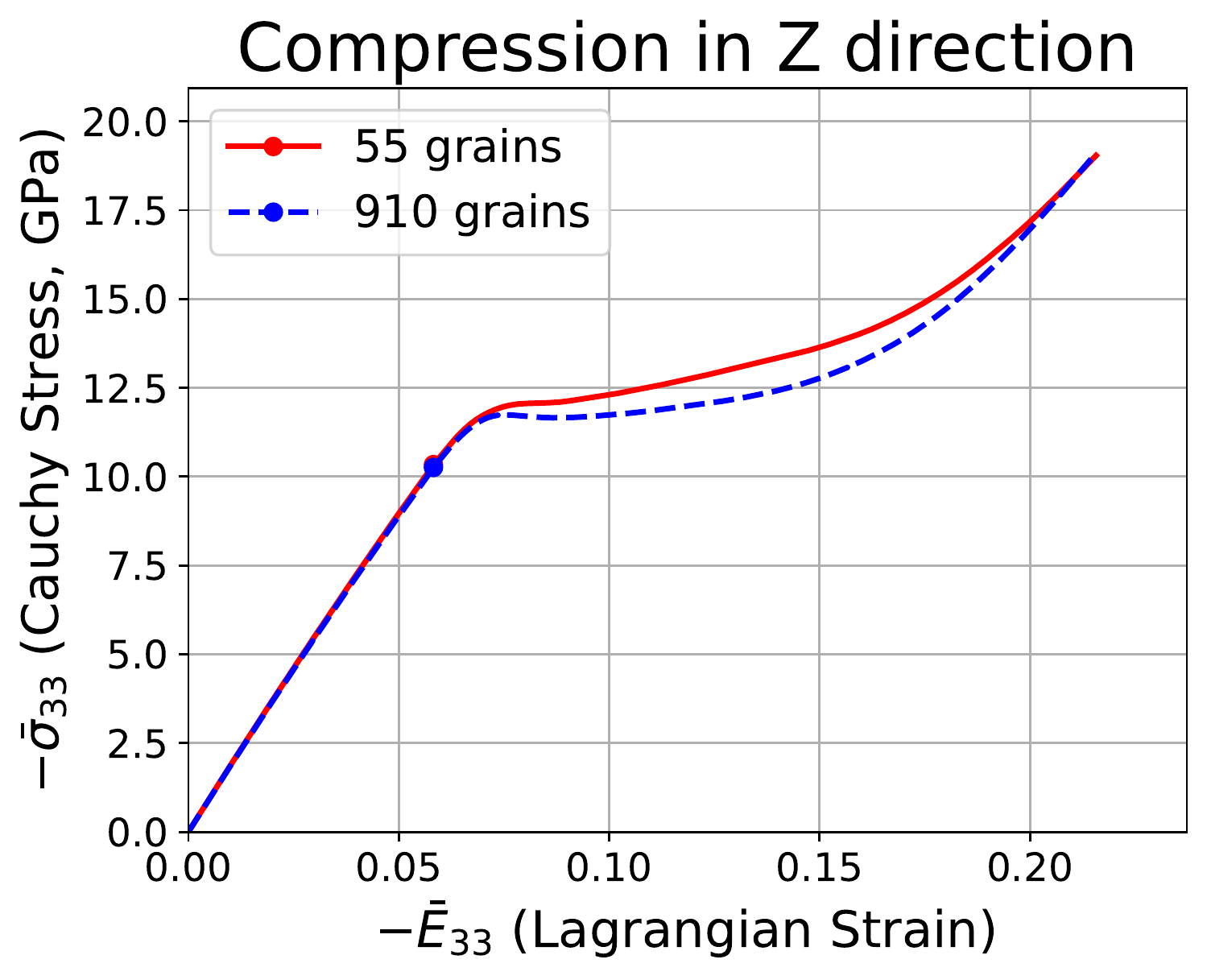}}
			\caption{Averaged Cauchy stress - Lagrangian strain plot for strain-controlled loading for 55 grains and 910 grains. The dot marker represents the onset of phase transformation.}
			\label{strain_con}
		\end{figure}
	\end{minipage}
	\hspace{0.05\linewidth}
	\begin{minipage}{0.45\linewidth}
		
	\end{minipage}
\end{minipage}
\hspace{5mm}

\begin{figure}
	\centering
	\resizebox{150mm}{!}{\includegraphics[trim={45mm 45mm 45mm 45mm}, clip, angle=270]{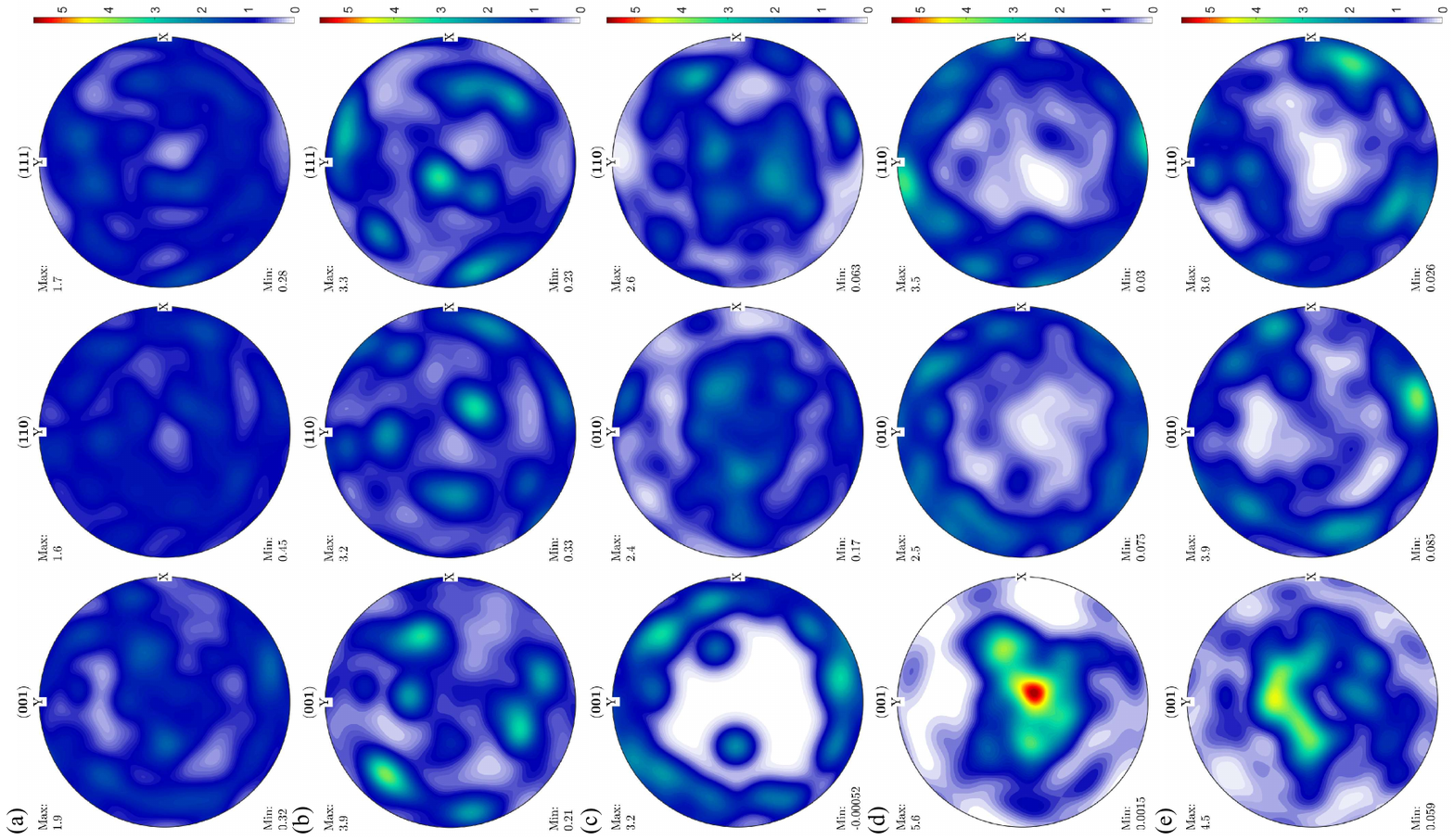}}
	\caption{Pole figures of (a) initial austenite, (b) residual austenite, (c) final martensitic  variant $M_1$,  (d) $M_2$, and (e) $M_3$ for strain-controlled loading of 55 grains. The minimum and maximum intensities for each direction are shown in the legend corresponding to each figure.}
	\label{pf_55}
\end{figure}

\begin{figure}
	\centering
	\resizebox{150mm}{!}{\includegraphics[trim={45mm 45mm 45mm 45mm}, clip, angle=270]{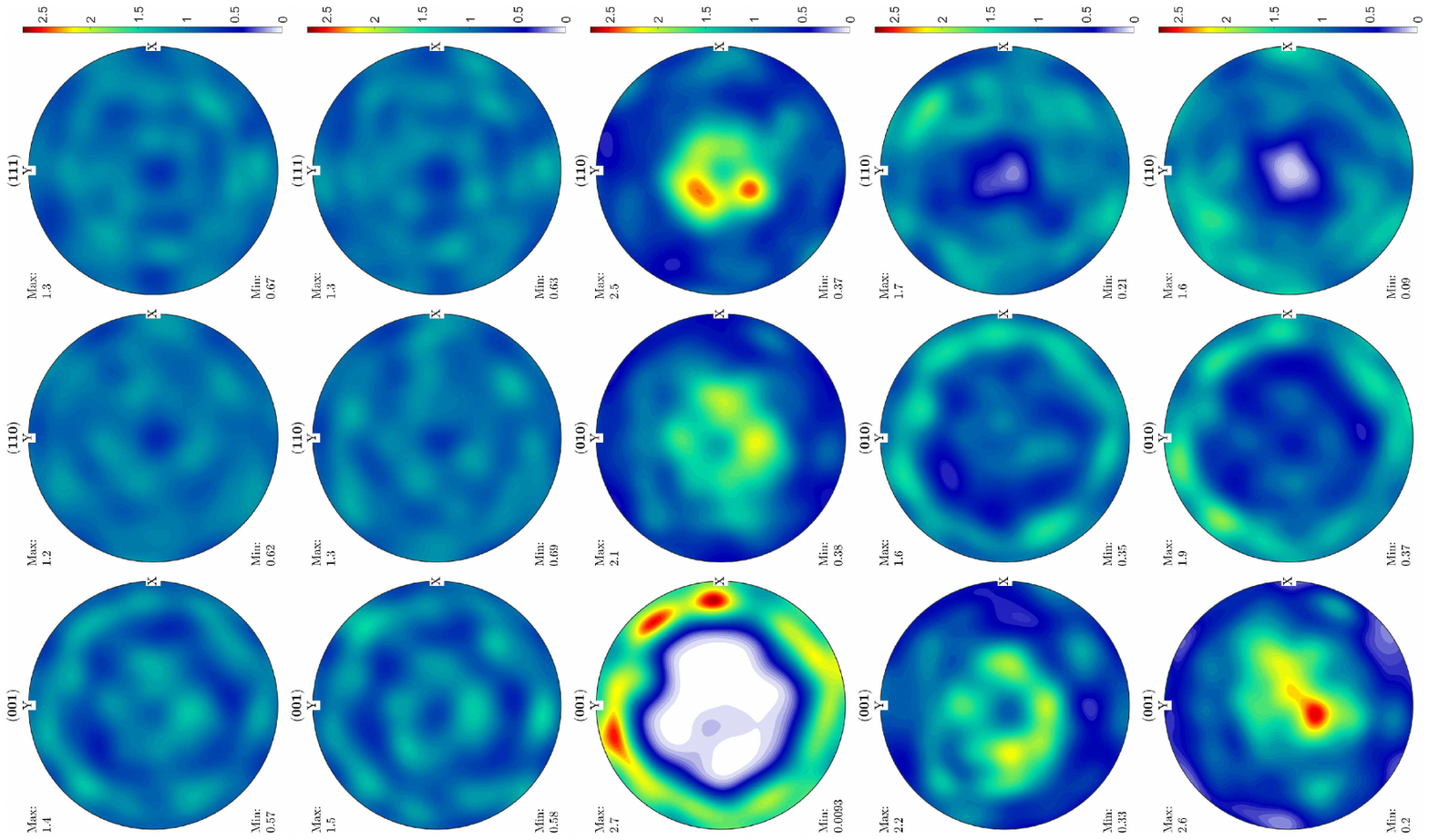}}
	\caption{Pole figures of (a) initial austenite, (b) residual austenite, (c) final martensitic  variant $M_1$,  (d) $M_2$, and (e) $M_3$ for strain-controlled loading of 55 grains. The minimum and maximum intensities for each direction are shown in the legend corresponding to each figure.}
	\label{pf_910}
\end{figure}

\Cref{pf_55,pf_910} shows the pole figures for 55 and 910 grains, respectively. \Cref{pf_55}(a) and \Cref{pf_910}(a) for initial austenite  demonstrate quite an even spread in all the directions because  the random  texture of the stress-free initial configuration was chosen. \Cref{pf_55}(a) shows few low-density regions because of the lower number of grains. Comparing the initial (\Cref{pf_55}(a)) and final (\Cref{pf_55}(b)) austenite for  55 grains,  we can clearly notice depletion of the austenite grains due to  their  transformation to martensite. Similar depletion is not noticed in \Cref{pf_910}(b) for 910 grains  because many grains do not  completely transform to martensite. The increase in density in \Cref{pf_55}(b) or \Cref{pf_910}(b) is because of the rotation of part of or entire austenitic grains, which was observed for single crystal as well \cite{babaei2020finite}.

\Cref{pf_55}(c)-(e) and \Cref{pf_910}(c)-(e) present the pole figures for the three martensitic variants at the last stage of simulations. As expected, there is a 90$^\circ$ rotation relation between the first and the second or the third variant about the $(001)$ axis. The trends in the volume fractions of the individual martensitic variant in \Cref{pf_55,pf_910} match that of \Cref{strain_vf}. The first variant is the lowest in both  cases, as only the grains that have their $c$-axis aligned with the loading direction can transform to the first variant. As for the second or the third variants, they primarily need to be oriented such that their $a$ or $b$-axis needs to be aligned with the loading direction. But because of the boundary conditions used, there are lateral stresses which can contribute to giving a resultant load in the preferred axes assisting in the transformation. This phenomenon is not possible for the first variant as the lateral stresses are only a fraction of the applied load and cannot be the primary contributor to fulfilling the phase transformation criteria.

\subsection{Stress-controlled loading}
\begin{figure}
	\centering
	\resizebox{150mm}{!}{\includegraphics[]{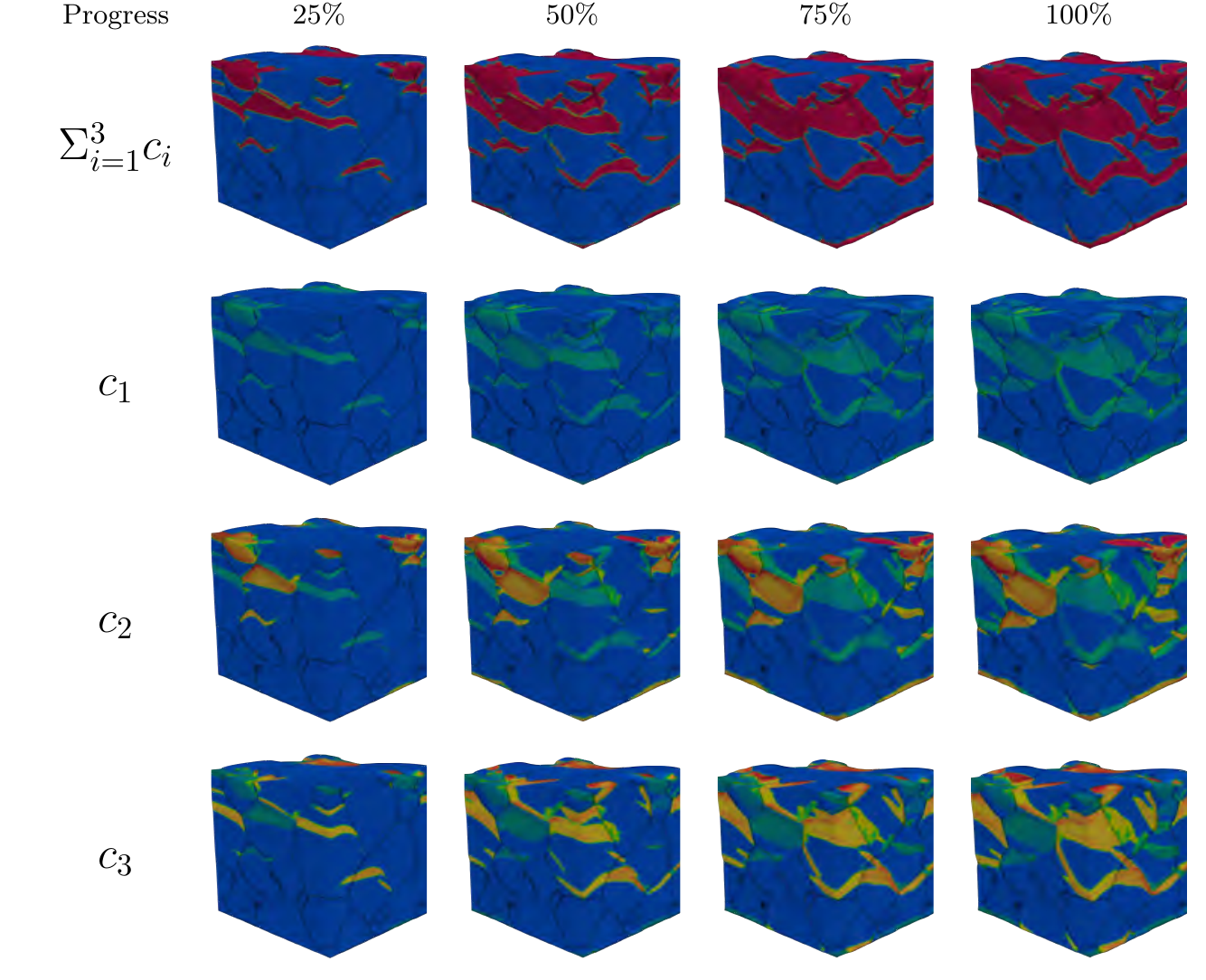}}
	\begin{minipage}{\linewidth}
	\centering
	\hspace{0.05\linewidth}
	\begin{minipage}{0.2\linewidth}
		\begin{figure}[H]
			\resizebox{20mm}{!}{\includegraphics{Images/axes.png}}
		\end{figure}
	\end{minipage}
	\hspace{0.05\linewidth}
	\begin{minipage}{0.5\linewidth}
		\begin{figure}[H]
			\resizebox{60mm}{!}{\includegraphics{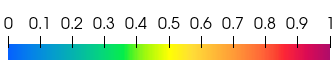}}
		\end{figure}
	\end{minipage}
\end{minipage}
	\caption{Evolution of volume fractions of phases for stress-controlled loading of a sample with 55 grains. }
	\label{55ss_cevol}
\end{figure}

\begin{figure}
	\centering
	\resizebox{150mm}{!}{\includegraphics[]{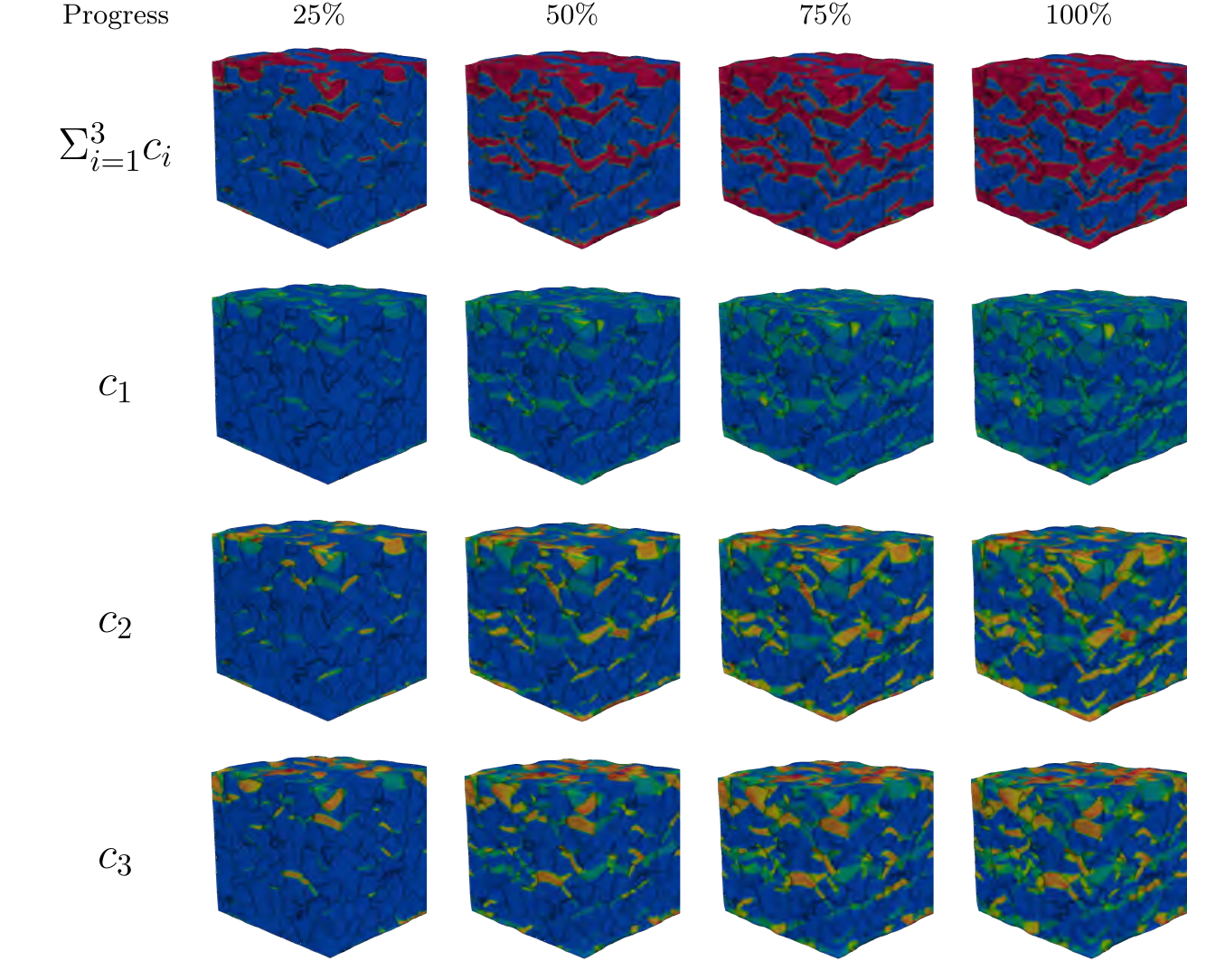}}
	\begin{minipage}{\linewidth}
	\centering
	\hspace{0.05\linewidth}
	\begin{minipage}{0.2\linewidth}
		\begin{figure}[H]
			\resizebox{20mm}{!}{\includegraphics{Images/axes.png}}
		\end{figure}
	\end{minipage}
	\hspace{0.05\linewidth}
	\begin{minipage}{0.5\linewidth}
		\begin{figure}[H]
			\resizebox{60mm}{!}{\includegraphics{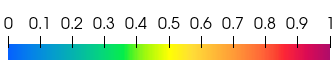}}
		\end{figure}
	\end{minipage}
\end{minipage}
	\caption{Evolution of volume fractions of phases for strain-controlled loading of a sample with  910 grains. }
	\label{910ss_cevol}
\end{figure}

\begin{figure}
	\centering
	\resizebox{150mm}{!}{\includegraphics[]{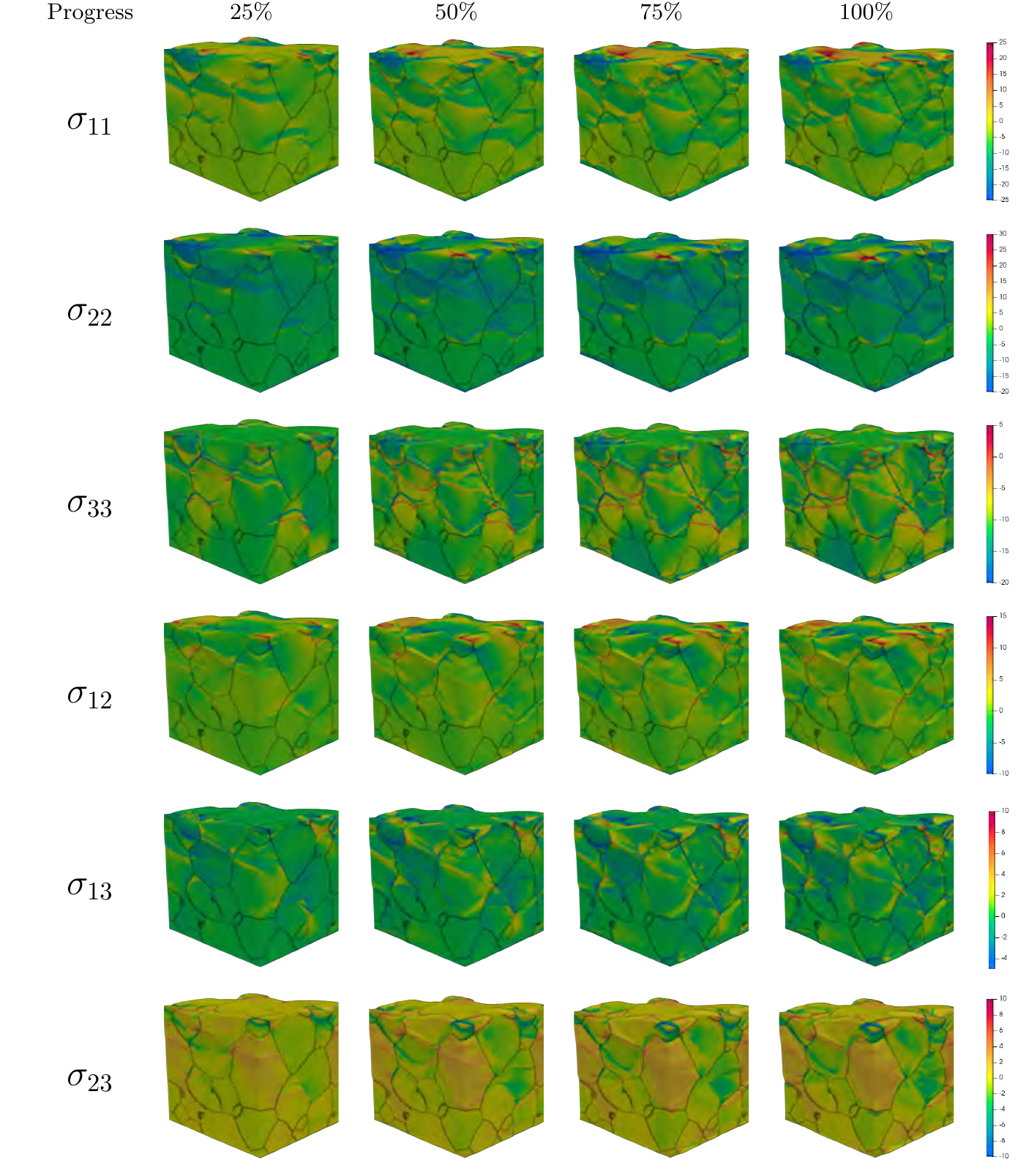}}
	\begin{minipage}{\linewidth}
	\hspace{0.1\linewidth}
	\begin{minipage}{0.1\linewidth}
		\begin{figure}[H]
			\resizebox{10mm}{!}{\includegraphics{Images/axes.png}}
		\end{figure}
	\end{minipage}
	\hspace{0.85\linewidth}
\end{minipage}
	\caption{Evolution of the components of the Cauchy stress tensor for stress-controlled loading of a sample with 55 grains.}
	\label{55ss_sevol}
\end{figure}

\begin{figure}
	\centering
	\resizebox{150mm}{!}{\includegraphics[]{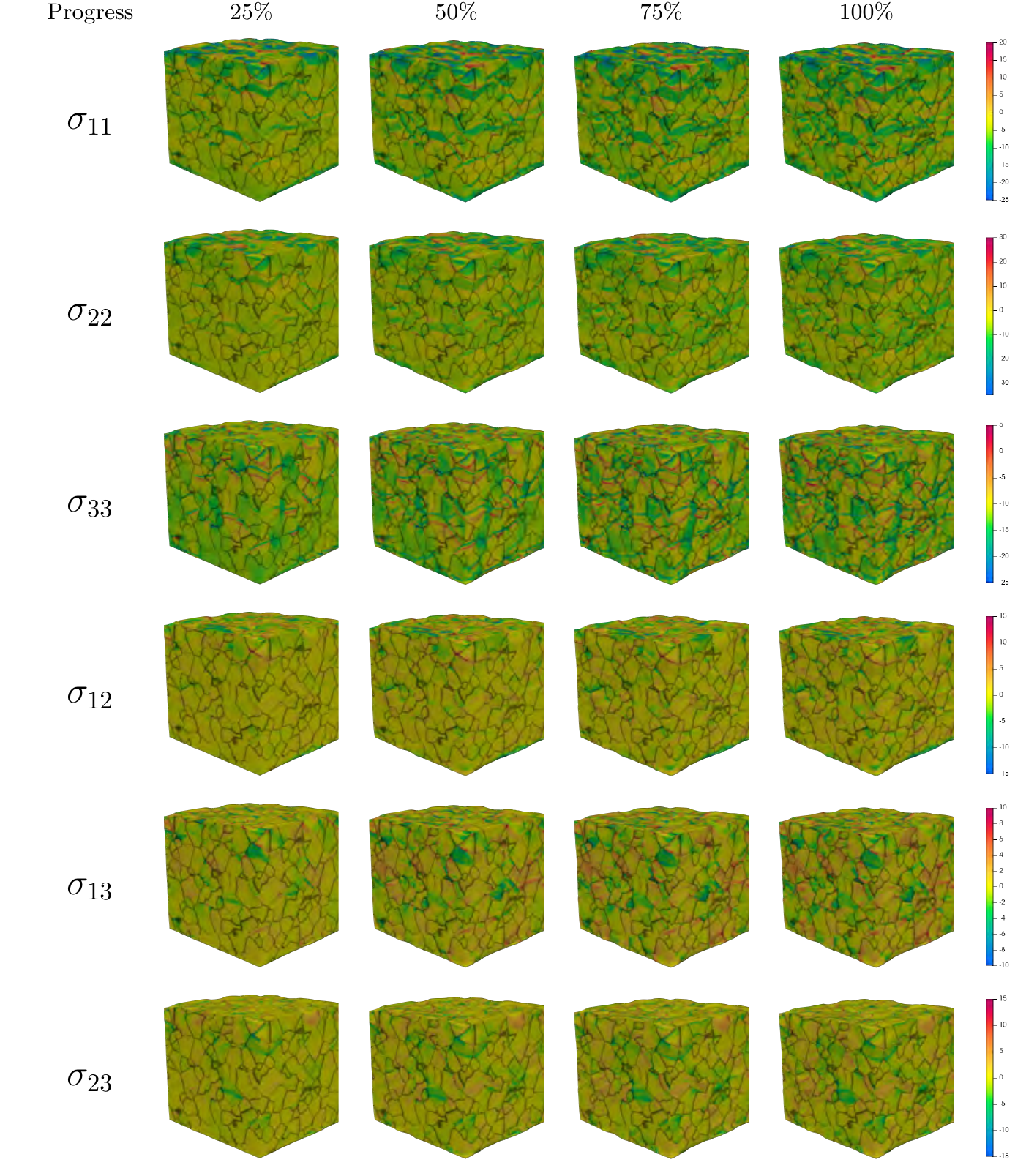}}
	\begin{minipage}{\linewidth}
	\hspace{0.1\linewidth}
	\begin{minipage}{0.1\linewidth}
		\begin{figure}[H]
			\resizebox{10mm}{!}{\includegraphics{Images/axes.png}}
		\end{figure}
	\end{minipage}
	\hspace{0.85\linewidth}
\end{minipage}
	\caption{Evolution of the components of the Cauchy stress tensor for strain-controlled loading of a sample with 910 grains.}
	\label{910ss_sevol}
\end{figure}

 In stress-controlled, the stress is applied at a rate of {-4 MPa/s till it reaches -11 GPa and held constant thereafter}. \Cref{55ss_cevol,910ss_cevol} and \cref{55ss_sevol,910ss_sevol} show the evolution of the volume fraction of Si II  and the individual martensitic variants as well as all stress components at different simulation stages for 55 and 910 grins. The videos showing the evolution of the volume fractions are provided in the supplementary material.
 Volume fraction of each martensitic variant $\bar{c}_i$ averaged over the sample based on Eq. \ref{av-c} and $\bar{c}$
vs. strain  is shown in Fig.	\ref{stress_vf}.
 There is no significant difference, in comparison with  the strain-controlled case, in nucleation at grain boundaries and triple junctions and character of growth of martensitic units, stress concentrations, and that $\bar{c}_2\simeq \bar{c}_3\simeq 2\bar{c}_1$.
 The same discussion on the luck of the physical sense in  $\bar{c}_i$ is valid;  $\bar{c}$  has a  physical meaning only.

\begin{minipage}{\linewidth}
	\centering
	\begin{minipage}{0.45\linewidth}
		\begin{figure}[H]
			\centering
			\resizebox{70mm}{!}{\includegraphics{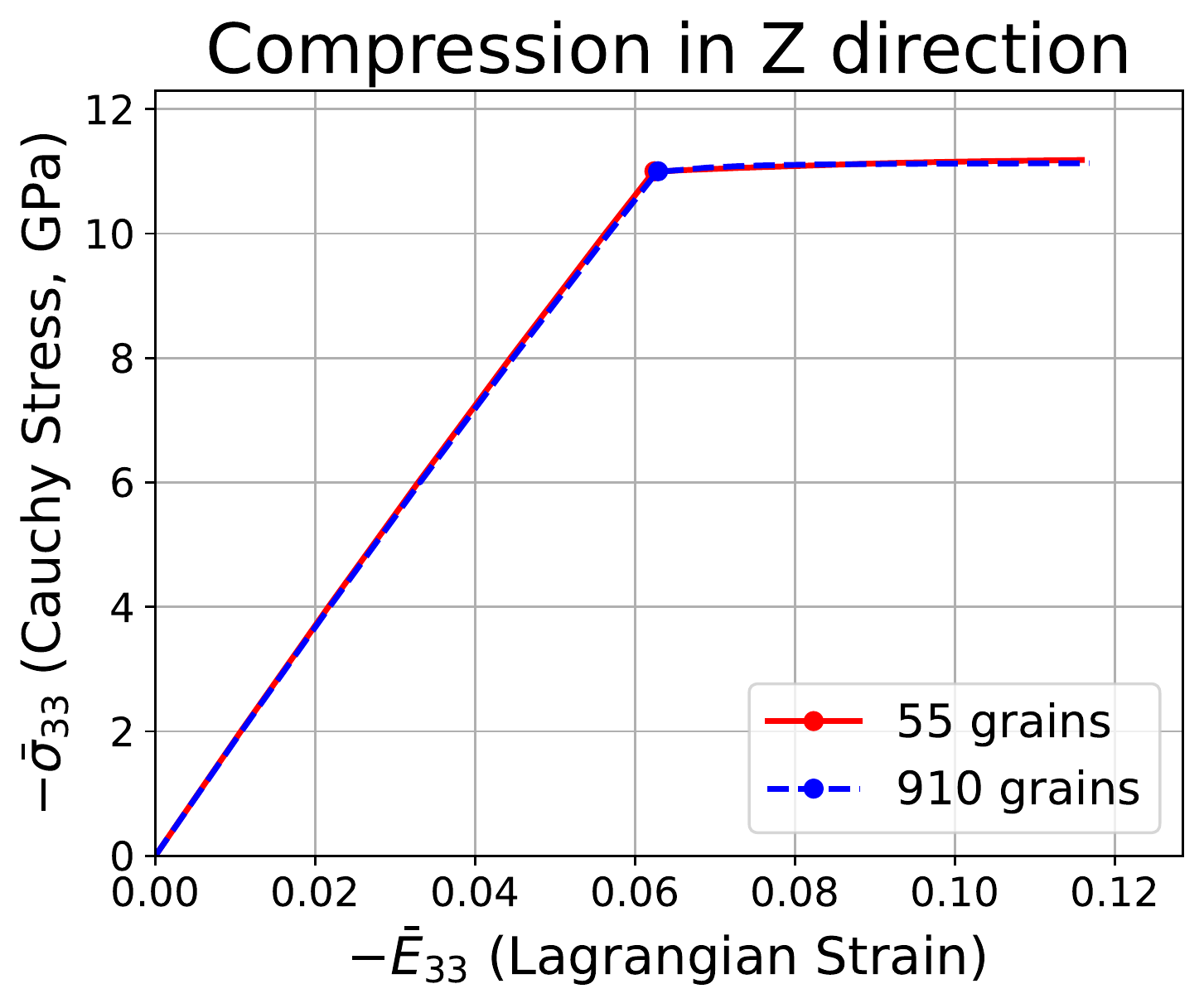}}
			\caption{True stress-strain plot for stress-controlled loading for 55  and 910 grains. The dot marker represents the onset of phase transformation.}
			\label{stress_con}
		\end{figure}
	\end{minipage}
	\hspace{0.05\linewidth}
	\begin{minipage}{0.45\linewidth}
		\begin{figure}[H]
			\centering
			\resizebox{70mm}{!}{\includegraphics{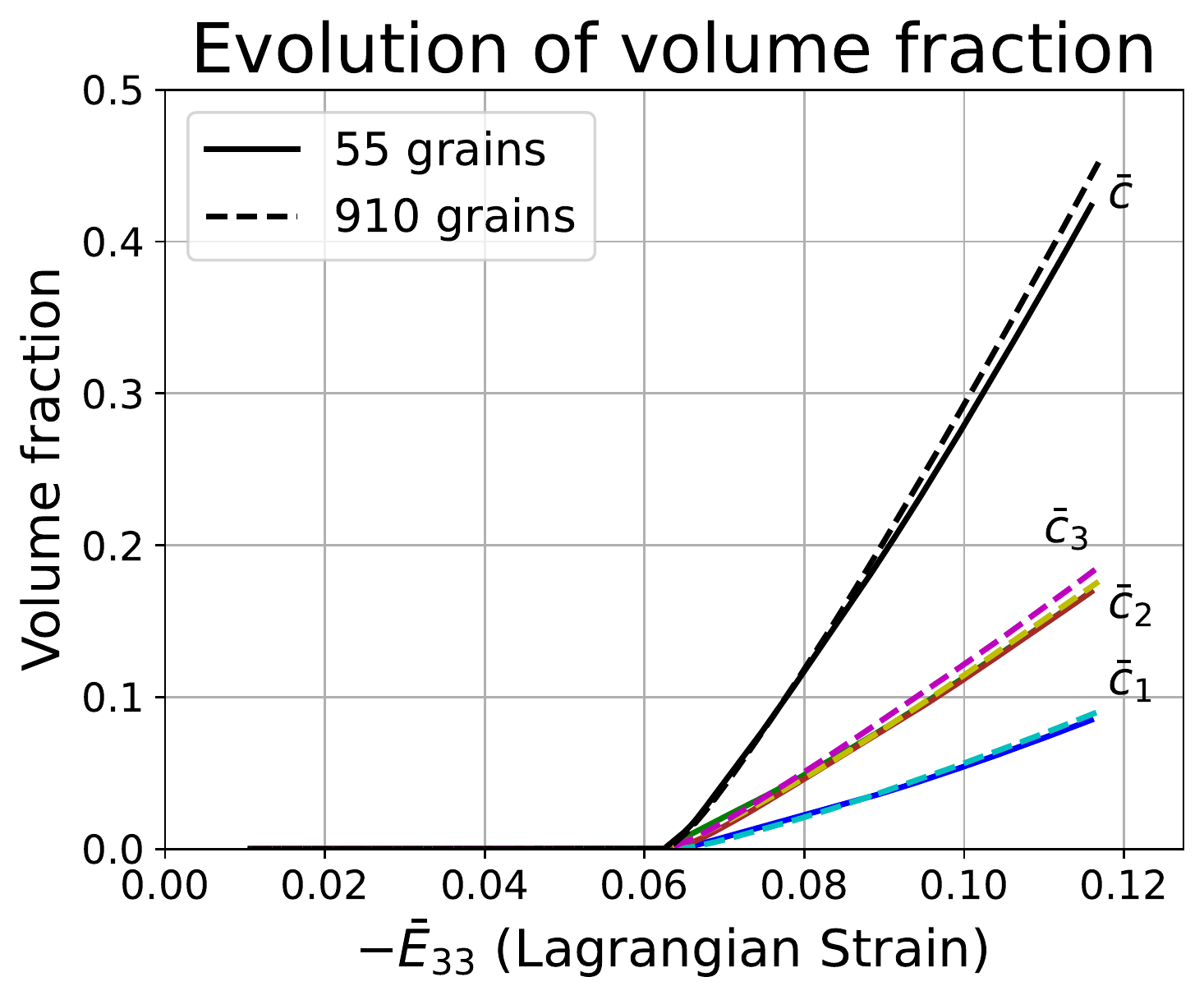}}
			\caption{Volume fraction of each martensitic variant $\bar{c}_i$ and total martensite $\bar{c}$) averaged over the sample based on Eq. \ref{av-c}
vs. strain  for stress-controlled loading.
The lack of physical sense for $\bar{c}_i$ is described in the text.}
			\label{stress_vf}
		\end{figure}
	\end{minipage}
\end{minipage}
\hspace{-5mm}

\begin{figure}
	\centering
	\resizebox{150mm}{!}{\includegraphics[]{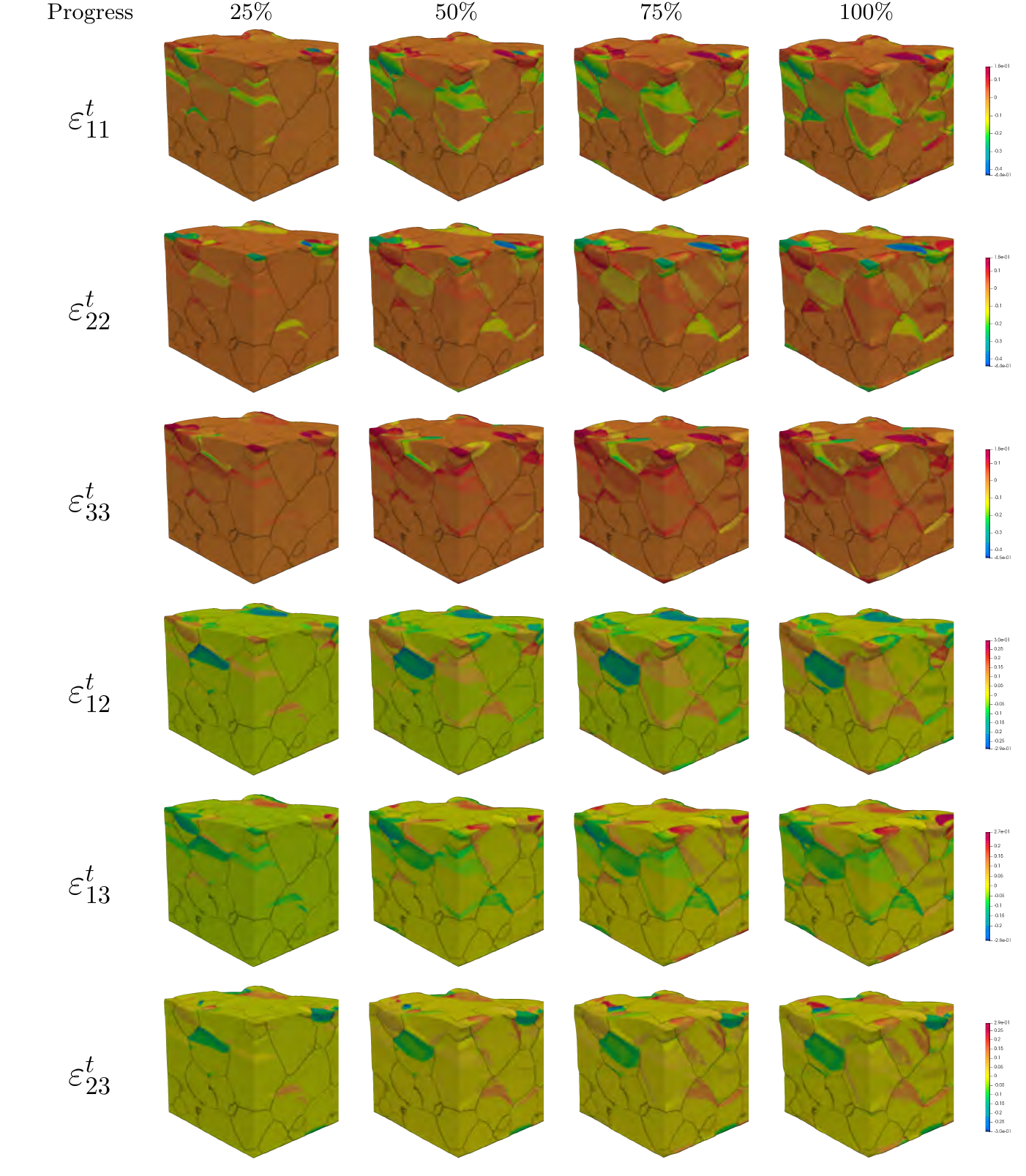}}
	\begin{minipage}{\linewidth}
	\hspace{0.1\linewidth}
	\begin{minipage}{0.1\linewidth}
		\begin{figure}[H]
			\resizebox{10mm}{!}{\includegraphics{Images/axes.png}}
		\end{figure}
	\end{minipage}
	\hspace{0.85\linewidth}
\end{minipage}
	\caption{Evolution of the components of the transformation strain tensor for stress-controlled loading of a sample with 55 grains.}
	\label{55ss_etevol}
\end{figure}

\begin{figure}
	\centering
	\resizebox{150mm}{!}{\includegraphics[]{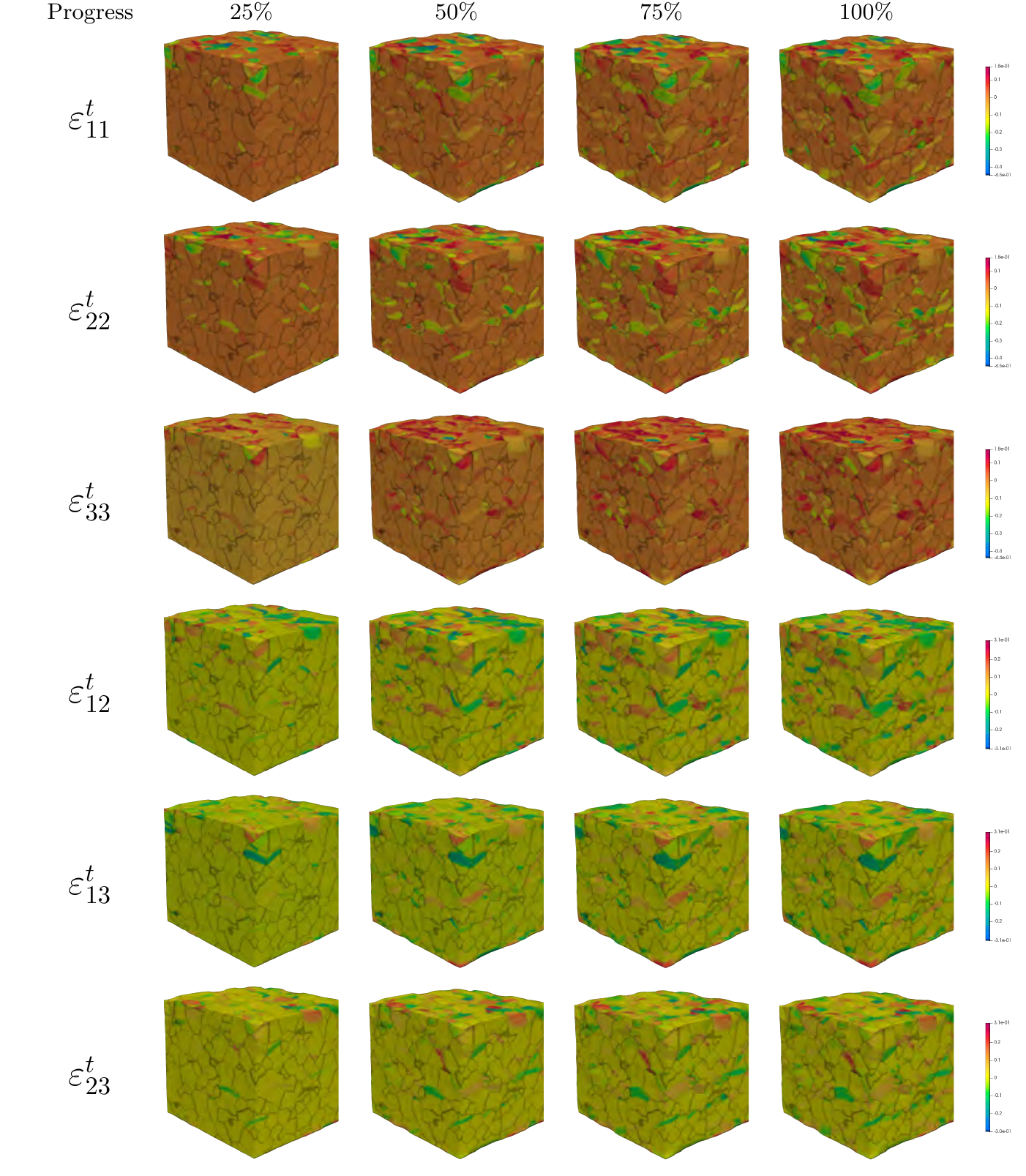}}
	\begin{minipage}{\linewidth}
	\hspace{0.1\linewidth}
	\begin{minipage}{0.1\linewidth}
		\begin{figure}[H]
			\resizebox{10mm}{!}{\includegraphics{Images/axes.png}}
		\end{figure}
	\end{minipage}
	\hspace{0.85\linewidth}
\end{minipage}
	\caption{Evolution of the components of the transformation strain tensor for strain-controlled loading of a sample with 910 grains.}
	\label{910ss_etevol}
\end{figure}

Stress-controlled compression for both  numbers of grains is up to  42\% transformation to Si II, after which process diverges. The stress-strain plots for strain-controlled loading are given in \Cref{stress_con}.  The PT starts immediately at -11 GPa and continues to the end of simulations at $E_{33}= -0.116$. These values are larger than
those for strain-controlled loading:  at -11 GPa, we have $E_{33}= -0.063$ and $c=2.4\times10^{-3}$ only; however, at $E_{33}=-0.116$ one gets
$c=0.362$ for strain-controlled loading.
 The possibility of larger strains and transformation progress at -11 GPa
is related to less constraint deformation at the horizontal external surfaces.
Periodic conditions for displacements along the axis 3 for strain-controlled loading lead to more homogenous PT near both horizontal surfaces
and small deviations from the flat surfaces. In contrast,  for stress-controlled loading, lack of periodic conditions  along the axis 3 leads to much pronounced PT near the upper horizontal surface, which spreads in the upper part of the sample.
This leads to the loss of stability of  Si II nuclei near stress concentrators, their fast growth, interaction,  coalescence, and more pronounced auto-catalytic effect.
Thus, specific boundary conditions are very influential, which is necessary to take into account in the problem formulation.
The difference in volume fractions of $M_i$ for 55 and 910 grains in \cref{strain_vf} and $\bar{c}$ is much smaller than for
 the strain-controlled loading in Fig.	\ref{strain_vf}, again due to less restrictive boundary conditions.

The peak stresses are large but smaller than for strain-controlled loading (\cref{55ss_sevol,910ss_sevol}). Thus, for small grains, $\sg_{22}$ varies from $- 33$ to $30$ GPa and $\sg_{33}$ varies from $- 25$ to $5$ GPa.
 Similar, shear stress $\sg_{12}$ and $\sg_{23}$ vary from $- 15$ to $15$ GPa.
For large grains, the magnitude of extremes in stresses are smaller by $5$ to $10$ GPa, like for strain-controlled loading.
Lower peak stresses are partially caused by a smaller volume fraction of Si II.

\begin{figure}[H]
	\centering
	\resizebox{150mm}{!}{\includegraphics[trim={45mm 45mm 45mm 45mm}, clip, angle=270]{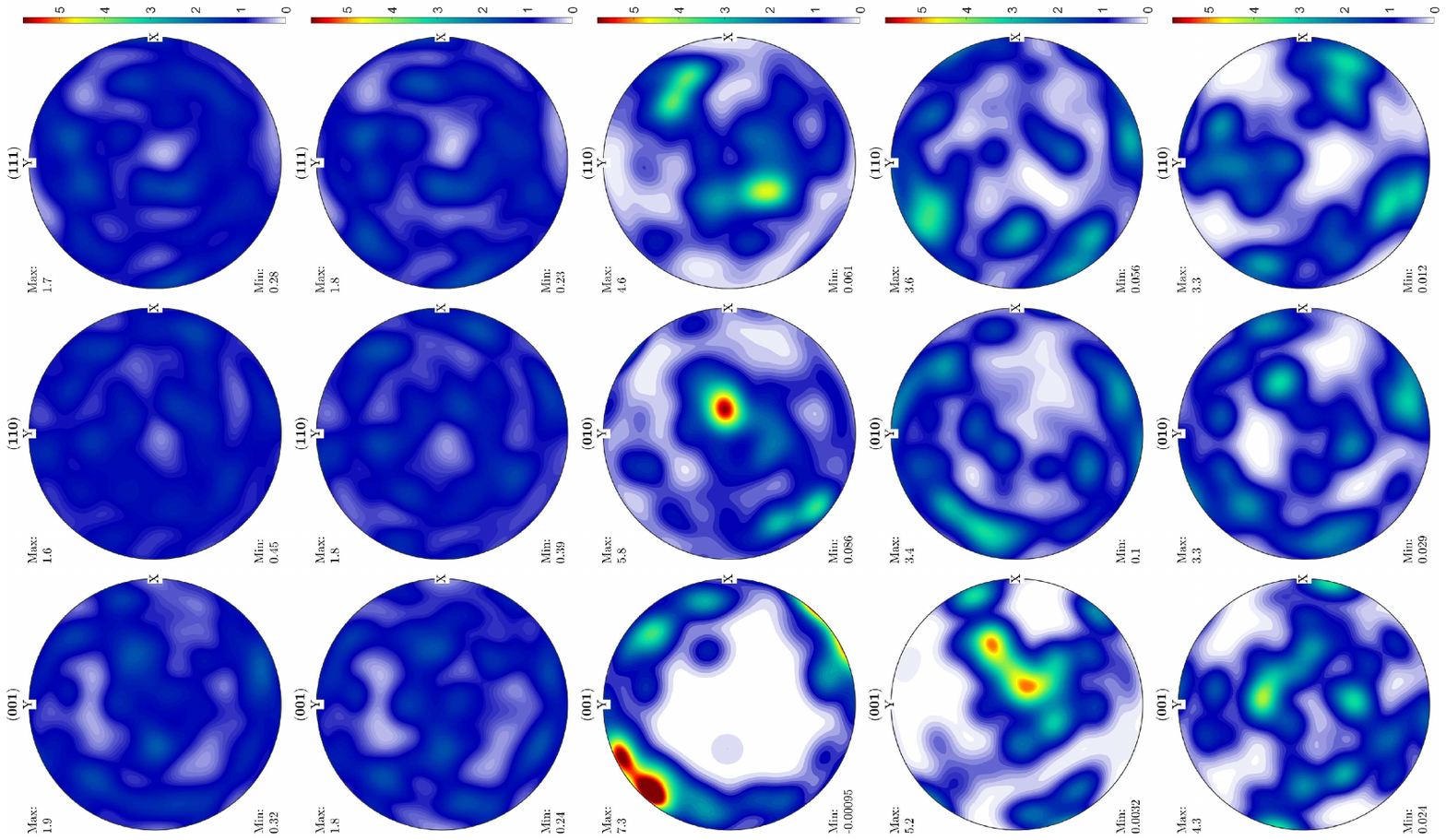}}
	\caption{Pole figures of (a) initial austenite, (b) residual austenite, (c) final martensitic  variant $M_1$,  (d) $M_2$, and (e) $M_3$ for stress-controlled loading of 55 grains. }
	\label{pf_55_ss}
\end{figure}

\begin{figure}[H]
	\centering
	\resizebox{150mm}{!}{\includegraphics[trim={45mm 45mm 45mm 45mm}, clip, angle=270]{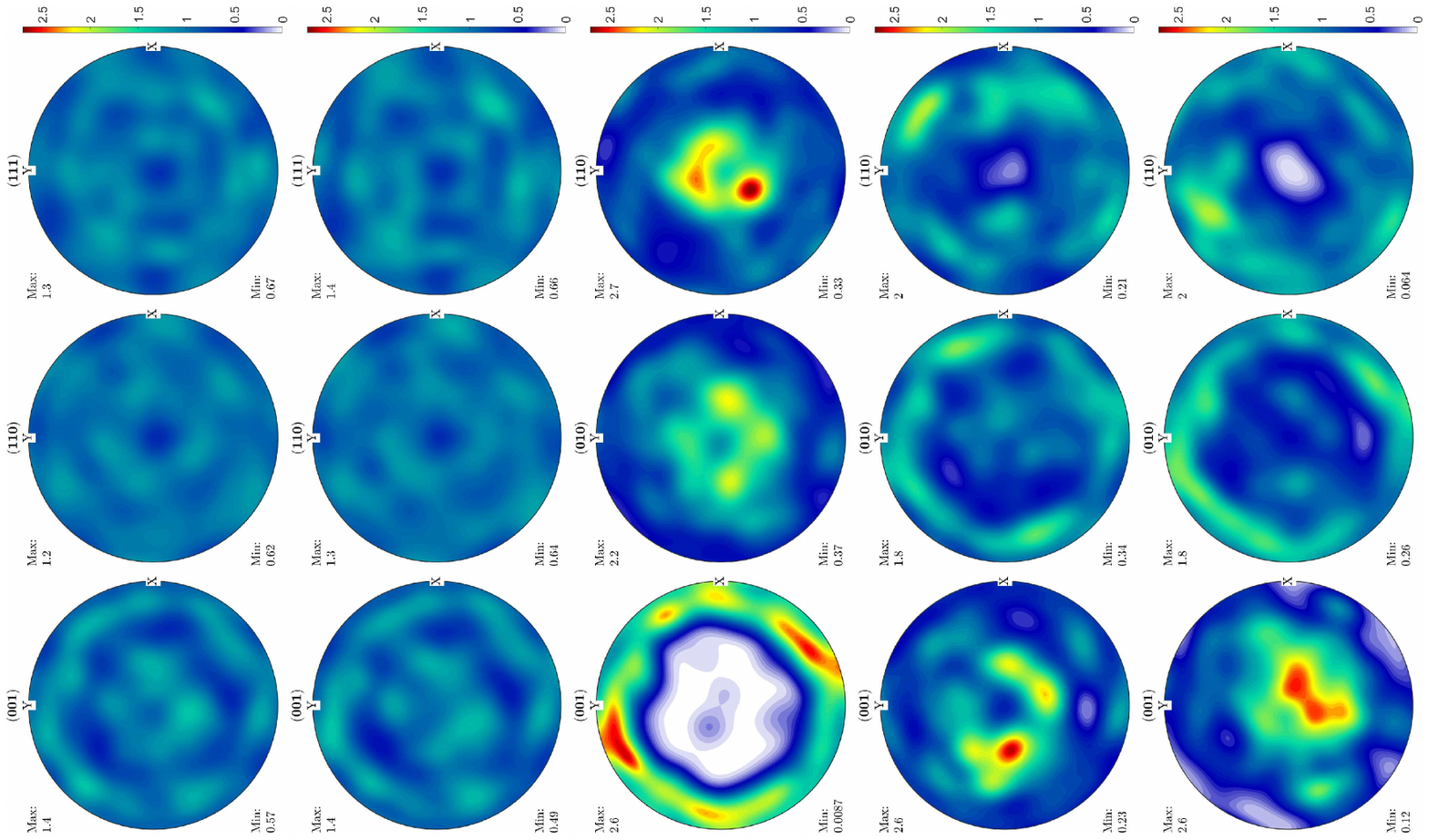}}
	\caption{Pole figures of (a) initial austenite, (b) residual austenite, (c) final martensitic  variant $M_1$,  (d) $M_2$, and (e) $M_3$ for stress-controlled loading of 910 grains. }
	\label{pf_910_ss}
\end{figure}

\Cref{pf_55_ss,pf_910_ss} show the pole figures for 55 and 910 grains, respectively, for the stress-controlled loading. \Cref{pf_55_ss}(a) and \Cref{pf_910_ss}(a) show initial austenite texture for both cases, which are the same as the ones chosen for strain-controlled loading.   For both 55 and 910  grains, there is no  significant depletion of the austenite, unlike for the strain-controlled loading, because the $\bar{c}=0.42$ only. This is reflected in the small difference between \cref{pf_55_ss}(a) and \cref{pf_55_ss}(b), and in sparsely distributed \cref{pf_55_ss}(c)-(e). Contrary to this, \cref{pf_910_ss}(c)-(e) show the pole figures with more uniform distributions because grains in many different orientations start transforming to martensite like in the case of \cref{pf_910}.

\section{Concluding remarks}\label{Conclusion}

In the paper, the first scale-free PFA modeling of the multivariant martensitic PT from cubic Si I to tetragonal Si II in a polycrystalline aggregate with up to 1000 grains is presented. All computational challenges related to large and very anisotropic transformation strain tensors,
the stress-tensor dependent athermal dissipative threshold for the
 PT, and potential elastic instabilities due to a variety of complex loadings in each grain, are overcome.
The importance of the simulations should also be stressed by the fact that since Si II does not exist under ambient conditions,
its microstructure cannot be studied using traditional post-mortem methods (SEM,  TEM, etc.).
For a single crystal, positions of Si I-Si II interfaces can be determined using in-situ high-pressure Laue diffraction, but still, in combination with molecular dynamics \cite{Chen2022NontrivialTransformation}. For polycrystals, this is currently impossible.

Coupled evolution of discrete martensitic microstructure, volume fractions of martensitic variants, and Si II,
 stress and transformation strain tensors, and texture are presented and analyzed.
  It is demonstrated   that the volume fraction of each martensitic variant ${c}_i$ in  each grain does not have a lot of sense unless the orientation of the grain is explicitly shown.
     The average over the sample $\bar{c}_i$ does not have any physical sense because the orientation of grains is not taken into account; they are misleading and cannot be used in the macroscopic theories.
 Macroscopic variables effectively
 representing multivariant   transformational behavior are introduced.
 One can present fields of  six components of
     $\fvep_{ti}c_i$   for each variant $M_i$ in the global coordinate system.
     More compact is  to
     present  six components of the total transformation strain
     $\fvep_{t}=\sum_{i=1}^m \fvep_{ti}c_i$  in the global coordinate system.
       For the averaged description, one can utilize six components of $\bar{\fvep}_{t}$.

 For strain-controlled uniaxial compression  with  periodic conditions in all directions,
 almost complete (96\%) PT was reached with small pockets of residual Si I.
 For stress-controlled  uniaxial compression  without  periodic conditions in the loading direction,
42\% of completion of was achieved at -11 GPa, much larger than 0.24\% for the same axial stress
   for strain-controlled loading, but relatively close volume fraction of Si II was reached for the same strain.
Lack of periodic conditions in the loading direction results in less constraint deformation at the horizontal external surfaces and more localized transformation near one of them, which leads to loss of stability of  Si II nuclei near stress concentrators, their fast growth, interaction, and coalescence.
Thus, tiny detail in the boundary conditions is very influential, which is necessary to take into account in the problem formulation.

In contrast to a single crystal, the local mechanical instabilities due to PT and negative local tangent modulus
are stabilized at the macroscale by arresting/slowing the growth of Si II regions by the grain boundaries and generating the internal back stresses. This leads to increasing magnitude of stress during PT.

Large transformation strains and grain boundaries lead
to huge internal stresses, which affect the microstructure evolution and macroscopic behavior.
The peak stresses reach 45 GPa in compression, 35 GPa in tension, and 20 GPa in shear for strain-controlled loading of 910 grains; for 55 grains, the magnitude of extremes in stresses are smaller by $5$ to $10$ GPa.
For stress-controlled loading and small grains,
the peak stresses reach 35 GPa in compression, 30 GPa in tension, and 15 GPa in shear;
for large grains, they are smaller by 5-10 GPa, like for strain-controlled loading.
Lower peak stresses for stress-controlled loading are partially caused by a smaller volume fraction of Si II.
Despite these differences, the macroscopic (overall) stress-strain and transformational behavior  for 55 and 910 grains are quite close and
differ less than by 10\% for strain-controlled loading and even less for stress-controlled one.  On the good side, this allows the determination of the macroscopic constitutive equations
by treating aggregate with a small number of grains.
On the bad side, that means that the current model does not describe experimentally
  observed effect of the grain size on the Si I to Si II PT pressure or stress \cite{Sorb2022,zeng2020origin,xuan2020pressure,tolbert1996pressure}.
  The reason is that the current model does not include dislocations as local stress concentrators in bulk and at grain boundaries. This will be the next step in developing the current model.
 We can use the same approach for introducing discrete dislocations via the solution of the contact problem, as it was done in \cite{Levitas2018Scale-FreeMicrostructure,Esfahani-etal-2020} for small strains and 2D formulations. Of course, it is much more challenging to do this for 3D and large strains. A more realistic model of the grain boundaries is required as well.
The developed methodology can be used for studying  various PTs with large transformation strains (e.g., hexagonal and rhombohedral graphite to hexagonal and cubic diamond, similar PTs from graphite-like BN to superhard diamond-like BN, PTs in semiconducting Ge and GaSb, etc.) and for further development for plastic strain-induced PTs.

\noindent {\bf Acknowledgement}
\par The support of NSF  (CMMI-1943710) and Iowa State University (Vance Coffman Faculty Chair  Professorship)  is	gratefully acknowledged. The simulations were performed at  Extreme Science and Engineering Discovery Environment (XSEDE), allocation TG-MSS170015.


\bibliography{ref_updated.bib}

\begin{thebibliography}{10}
\expandafter\ifx\csname url\endcsname\relax
  \def\url#1{\texttt{#1}}\fi
\expandafter\ifx\csname urlprefix\endcsname\relax\def\urlprefix{URL }\fi
\expandafter\ifx\csname href\endcsname\relax
  \def\href#1#2{#2} \def\path#1{#1}\fi

\bibitem{heath2020research}
G.~A. Heath, T.~J. Silverman, M.~Kempe, M.~Deceglie, D.~Ravikumar, T.~Remo,
  H.~Cui, P.~Sinha, C.~Libby, S.~Shaw, et~al., {Research and development
  priorities for silicon photovoltaic module recycling to support a circular
  economy}, Nature Energy 5~(7) (2020) 502--510.

\bibitem{chelikowsky2004introduction}
J.~Chelikowsky, Introduction: Silicon in all its forms, in: P.~Siffert, E.~F.
  Krimmel (Eds.), Silicon: Evolution and Future of a Technology, Springer
  Berlin Heidelberg, Berlin, Heidelberg, 2004, pp. 1--22.

\bibitem{Domnich-2004}
V.~Domnich, D.~Ge, Y.~Gogotsi, {Indentation-induced phase transformations in
  semiconductors}, in: V.~Domnich, Y.~Gogotsi (Eds.), High Pressure Surface
  Science and Engineering; Section 5.1, Institute of Physics, Bristol, 2004,
  pp. 381--442.

\bibitem{Kiran2015NanoindentationGermanium}
M.~S. Kiran, B.~Haberl, J.~E. Bradby, J.~S. Williams, {Chapter Five -
  Nanoindentation of Silicon and Germanium}, in: L.~Romano, V.~Privitera,
  C.~Jagadish (Eds.), Defects in Semiconductors, Vol.~91 of Semiconductors and
  Semimetals, Elsevier, 2015, pp. 165--203.
\newblock \href {https://doi.org/https://doi.org/10.1016/bs.semsem.2014.12.002}
  {\path{doi:https://doi.org/10.1016/bs.semsem.2014.12.002}}.

\bibitem{Goel2015DiamondSimulation}
S.~Goel, X.~Luo, A.~Agrawal, R.~L. Reuben, {Diamond machining of silicon: a
  review of advances in molecular dynamics simulation}, International Journal
  of Machine Tools and Manufacture 88 (2015) 131--164.

\bibitem{Ikoma2012PhaseTorsion}
Y.~Ikoma, K.~Hayano, K.~Edalati, K.~Saito, Q.~Guo, Z.~Horita, {Phase
  transformation and nanograin refinement of silicon by processing through
  high-pressure torsion}, Applied Physics Letters 101~(12) (2012) 121908.

\bibitem{Ikoma2014FabricationTorsion}
Y.~Ikoma, K.~Hayano, K.~Edalati, K.~Saito, Q.~Guo, Z.~Horita, T.~Aoki, D.~J.
  Smith, {Fabrication of nanograined silicon by high-pressure torsion}, Journal
  of Materials Science 49 (2014) 6565--6569.

\bibitem{patten2019ductile}
J.~A. Patten, H.~Cherukuri, J.~Yan, {Ductile-regime machining of semiconductors
  and ceramics}, in: V.~Domnich, Y.~Gogotsi (Eds.), High-Pressure Surface
  Science and Engineering, Institute of Physics, Bristol, 2004, pp. 543--632.

\bibitem{malyushitskaya1999mechanisms}
Z.~Malyushitskaya, {Mechanisms responsible for the strain-induced formation of
  metastable high-pressure Si, Ge, and GaSb phases with distorted tetrahedral
  coordination}, Inorganic Materials 35~(5) (1999) 425--430.

\bibitem{Levitas2017LatticeCriterion}
V.~I. Levitas, H.~Chen, L.~Xiong, {Lattice instability during phase
  transformations under multiaxial stress: Modified transformation work
  criterion}, Phys. Rev. B 96 (2017) 054118.
\newblock \href {https://doi.org/10.1103/PhysRevB.96.054118}
  {\path{doi:10.1103/PhysRevB.96.054118}}.

\bibitem{Levitas2017Triaxial-Stress-InducedPhases}
V.~I. Levitas, H.~Chen, L.~Xiong, {Triaxial-Stress-Induced Homogeneous
  Hysteresis-Free First-Order Phase Transformations with Stable Intermediate
  Phases}, Phys. Rev. Lett. 118 (2017) 025701.
\newblock \href {https://doi.org/10.1103/PhysRevLett.118.025701}
  {\path{doi:10.1103/PhysRevLett.118.025701}}.

\bibitem{Levitas-PRB-2004}
V.~I. Levitas, {High-pressure mechanochemistry: Conceptual multiscale theory
  and interpretation of experiments}, Phys. Rev. B 70 (2004) 184118.
\newblock \href {https://doi.org/10.1103/PhysRevB.70.184118}
  {\path{doi:10.1103/PhysRevB.70.184118}}.

\bibitem{levitas2018high}
V.~I. Levitas, {High pressure phase transformations revisited}, Journal of
  Physics: Condensed Matter 30~(16) (2018) 163001.
\newblock \href {https://doi.org/10.1088/1361-648X/aab4b0}
  {\path{doi:10.1088/1361-648X/aab4b0}}.

\bibitem{Levitas2019High-pressureAnvils}
V.~I. Levitas, {High-pressure phase transformations under severe plastic
  deformation by torsion in rotational anvils}, Materials Transactions 60~(7)
  (2019) 1294--1301.

\bibitem{Blank2013}
V.~D. Blank, E.~I. Estrin, Phase transitions in solids under high pressure, CRC
  Press, 2013.

\bibitem{Bridgman1935EffectsPressure}
P.~W. Bridgman, {Effects of High Shearing Stress Combined with High Hydrostatic
  Pressure}, Phys. Rev. 48 (1935) 825--847.
\newblock \href {https://doi.org/10.1103/PhysRev.48.825}
  {\path{doi:10.1103/PhysRev.48.825}}.

\bibitem{edalati2016review}
K.~Edalati, Z.~Horita, {A review on high-pressure torsion (HPT) from 1935 to
  1988}, Materials Science and Engineering: A 652 (2016) 325--352.
\newblock \href {https://doi.org/https://doi.org/10.1016/j.msea.2015.11.074}
  {\path{doi:https://doi.org/10.1016/j.msea.2015.11.074}}.

\bibitem{Gao2019}
Y.~Gao, Y.~Ma, Q.~An, V.~Levitas, Y.~Zhang, B.~Feng, J.~Chaudhuri, W.~A.
  Goddard, {Shear driven formation of nano-diamonds at sub-gigapascals and 300
  K}, Carbon 146 (2019) 364--368.
\newblock \href {https://doi.org/https://doi.org/10.1016/j.carbon.2019.02.012}
  {\path{doi:https://doi.org/10.1016/j.carbon.2019.02.012}}.

\bibitem{Levitas2002Low-pressureTheory}
V.~I. Levitas, L.~K. Shvedov, {Low-pressure phase transformation from
  rhombohedral to cubic BN: Experiment and theory}, Phys. Rev. B 65 (2002)
  104109.
\newblock \href {https://doi.org/10.1103/PhysRevB.65.104109}
  {\path{doi:10.1103/PhysRevB.65.104109}}.

\bibitem{ji2012shear}
C.~Ji, V.~I. Levitas, H.~Zhu, J.~Chaudhuri, A.~Marathe, Y.~Ma, {Shear-induced
  phase transition of nanocrystalline hexagonal boron nitride to wurtzitic
  structure at room temperature and lower pressure}, Proceedings of the
  National Academy of Sciences 109~(47) (2012) 19108--19112.

\bibitem{Pandey-Levitas-2020}
K.~Pandey, V.~I. Levitas, {In situ quantitative study of plastic strain-induced
  phase transformations under high pressure: Example for ultra-pure Zr}, Acta
  Materialia 196 (2020) 338--346.
\newblock \href {https://doi.org/https://doi.org/10.1016/j.actamat.2020.06.015}
  {\path{doi:https://doi.org/10.1016/j.actamat.2020.06.015}}.

\bibitem{levitas2022laws}
V.~Levitas, F.~Lin, K.~Pandey, S.~Yesudhas, C.~Park, {Laws of high-pressure
  phase and nanostructure evolution and severe plastic flow, {\it September
  9}}, Research Square (2022) 29\href
  {https://doi.org/https://doi.org/10.21203/rs.3.rs-1998605/v1}
  {\path{doi:https://doi.org/10.21203/rs.3.rs-1998605/v1}}.

\bibitem{aleksandrova1993phase}
M.~Aleksandrova, V.~Blank, S.~Buga, {Phase transitions in Ge and Si under shear
  deformation at pressure up to 12 GPa conditions and P-T- gamma[shear]
  diagrams of these elements}, Physics of the Solid State 35~(5) (1993)
  1308--17.

\bibitem{Sorb2022}
S.~Yesudhas, V.~Levitas, F.~Lin, K.~Pandey, J.~Smith, {Plastic strain induced
  phase transformations in micron and nano silicon}, {\it in preparation}
  (2023).

\bibitem{Voronin-etal-2003}
G.~Voronin, C.~Pantea, T.~Zerda, L.~Wang, Y.~Zhao, {In situ x-ray diffraction
  study of silicon at pressures up to 15.5 GPa and temperatures up to 1073 K},
  Physical Review B 68~(2) (2003) 020102.

\bibitem{Pokluda-etal-2015}
J.~Pokluda, M.~{\v{C}}ern{\`y}, M.~{\v{S}}ob, Y.~Umeno, {Ab initio calculations
  of mechanical properties: Methods and applications}, Progress in Materials
  Science 73 (2015) 127--158.

\bibitem{Umeno-Cerny-PRB-08}
Y.~Umeno, M.~{\v{C}}ern{\`y}, {Effect of normal stress on the ideal shear
  strength in covalent crystals}, Physical Review B 77~(10) (2008) 100101.

\bibitem{Telyatniketal-16}
R.~Telyatnik, A.~Osipov, S.~Kukushkin, {Ab initio modelling of nonlinear
  elastoplastic properties of diamond-like C, SiC, Si, Ge crystals upon large
  strains.}, Materials Physics \& Mechanics 29~(1) (2016) 1--16.

\bibitem{Cernyetal-JPCM-13}
M.~{\v{C}}ern{\`y}, P.~{\v{R}}eh{\'a}k, Y.~Umeno, J.~Pokluda, {Stability and
  strength of covalent crystals under uniaxial and triaxial loading from first
  principles}, Journal of Physics: Condensed Matter 25~(3) (2012) 035401.

\bibitem{zarkevich2018lattice}
N.~A. Zarkevich, H.~Chen, V.~I. Levitas, D.~D. Johnson, {Lattice instability
  during solid-solid structural transformations under a general applied stress
  tensor: Example of Si I $\rightarrow$ Si II with metallization}, Physical
  Review Letters 121~(16) (2018) 165701.

\bibitem{Chen-etal-NPJ-2020}
H.~Chen, N.~A. Zarkevich, V.~I. Levitas, D.~D. Johnson, X.~Zhang, {Fifth-degree
  elastic energy for predictive continuum stress--strain relations and elastic
  instabilities under large strain and complex loading in silicon}, NPJ
  Computational Materials 6~(1) (2020) 115.

\bibitem{Levitas2013Phase-fieldStrains}
V.~I. Levitas, {Phase-field theory for martensitic phase transformations at
  large strains}, International Journal of Plasticity 49 (2013) 85--118.

\bibitem{valentini2007phase}
P.~Valentini, W.~W. Gerberich, T.~Dumitric\ifmmode~\u{a}\else \u{a}\fi{},
  {Phase-Transition Plasticity Response in Uniaxially Compressed Silicon
  Nanospheres}, Phys. Rev. Lett. 99 (2007) 175701.
\newblock \href {https://doi.org/10.1103/PhysRevLett.99.175701}
  {\path{doi:10.1103/PhysRevLett.99.175701}}.

\bibitem{chrobak2011deconfinement}
D.~Chrobak, N.~Tymiak, A.~Beaber, O.~Ugurlu, W.~W. Gerberich, R.~Nowak,
  {Deconfinement leads to changes in the nanoscale plasticity of silicon},
  Nature Nanotechnology 6~(8) (2011) 480--484.

\bibitem{Chen_2019}
H.~Chen, V.~I. Levitas, L.~Xiong, {Amorphization induced by 60$^\circ$ shuffle
  dislocation pileup against different grain boundaries in silicon bicrystal
  under shear}, Acta Materialia 179 (2019) 287--295.

\bibitem{Zhang2019MolecularSilicon}
L.~C. Zhang, W.~C.~D. Cheong, {Molecular dynamics simulation of phase
  transformations in monocrystalline silicon}, in: V.~Domnich, Y.~Gogotsi
  (Eds.), High Pressure Surface Science and Engineering, Institute of Physics,
  Bristol, 2004, pp. 57--119.

\bibitem{Chen2022NontrivialTransformation}
H.~Chen, V.~I. Levitas, D.~Popov, N.~Velisavljevic, {Nontrivial nanostructure,
  stress relaxation mechanisms, and crystallography for pressure-induced Si-I
  $\rightarrow$ Si-II phase transformation}, Nature Communications 13~(1)
  (2022) 982.
\newblock \href {https://doi.org/10.1038/S41467-022-28604-1}
  {\path{doi:10.1038/S41467-022-28604-1}}.

\bibitem{Levitas2018PhaseTheory}
V.~I. Levitas, {Phase field approach for stress- and temperature-induced phase
  transformations that satisfies lattice instability conditions. Part I.
  General theory}, International Journal of Plasticity 106 (2018) 164--185.
\newblock \href {https://doi.org/10.1016/J.IJPLAS.2018.03.007}
  {\path{doi:10.1016/J.IJPLAS.2018.03.007}}.

\bibitem{Babaei_2018}
H.~Babaei, V.~I. Levitas, {Phase-field approach for stress- and
  temperature-induced phase transformations that satisfies lattice instability
  conditions. Part 2. Simulations of phase transformations Si I
  $\leftrightarrow$ Si II}, International Journal of Plasticity 107 (2018)
  223--245.
\newblock \href {https://doi.org/10.1016/j.ijplas.2018.04.006}
  {\path{doi:10.1016/j.ijplas.2018.04.006}}.

\bibitem{Babaei-Levitas-ActaMat-2019}
H.~Babaei, V.~I. Levitas, {Effect of 60$^\circ$ dislocation on transformation
  stresses, nucleation, and growth for phase transformations between silicon I
  and silicon II under triaxial loading: Phase-field study}, Acta Materialia
  177 (2019) 178--186.

\bibitem{Babaei2020Stress-MeasureTransformations}
H.~Babaei, V.~I. Levitas, {Stress-Measure Dependence of Phase Transformation
  Criterion under Finite Strains: Hierarchy of Crystal Lattice Instabilities
  for Homogeneous and Heterogeneous Transformations}, Phys. Rev. Lett. 124
  (2020) 075701.
\newblock \href {https://doi.org/10.1103/PhysRevLett.124.075701}
  {\path{doi:10.1103/PhysRevLett.124.075701}}.

\bibitem{Babaei2019AlgorithmicStrains}
H.~Babaei, A.~Basak, V.~I. Levitas, {Algorithmic aspects and finite element
  solutions for advanced phase field approach to martensitic phase
  transformation under large strains}, Computational Mechanics 64~(4) (2019)
  1177--1197.
\newblock \href {https://doi.org/10.1007/S00466-019-01699-Y}
  {\path{doi:10.1007/S00466-019-01699-Y}}.

\bibitem{Levitas2011Phase-fieldEnergy}
V.~I. Levitas, M.~Javanbakht, {Phase-field approach to martensitic phase
  transformations: Effect of martensite-martensite interface energy},
  International Journal of Materials Research 102~(6) (2011) 652--665.
\newblock \href {https://doi.org/10.3139/146.110529}
  {\path{doi:10.3139/146.110529}}.

\bibitem{Levitas-etal-2004}
V.~I. Levitas, A.~V. Idesman, D.~L. Preston, {Microscale simulation of
  martensitic microstructure evolution}, Physical Review Letters 93~(10) (2004)
  105701.

\bibitem{Idesman-etal-2005}
A.~V. Idesman, V.~I. Levitas, D.~L. Preston, J.-Y. Cho, {Finite element
  simulations of martensitic phase transitions and microstructures based on a
  strain softening model}, Journal of the Mechanics and Physics of Solids
  53~(3) (2005) 495--523.

\bibitem{Esfahani-etal-2018}
S.~E. Esfahani, I.~Ghamarian, V.~I. Levitas, P.~C. Collins, {Microscale phase
  field modeling of the martensitic transformation during cyclic loading of
  NiTi single crystal}, International Journal of Solids and Structures 146
  (2018) 80--96.

\bibitem{babaei2020finite}
H.~Babaei, V.~I. Levitas, {Finite-strain scale-free phase-field approach to
  multivariant martensitic phase transformations with stress-dependent
  effective thresholds}, Journal of the Mechanics and Physics of Solids 144
  (2020) 104114.

\bibitem{Levitas-Mahdi-Nanoscale-14}
V.~I. Levitas, M.~Javanbakht, {Phase transformations in nanograin materials
  under high pressure and plastic shear: nanoscale mechanisms}, Nanoscale 6~(1)
  (2014) 162--166.

\bibitem{Javanbakht2018NanoscaleStudy}
M.~Javanbakht, V.~I. Levitas, {Nanoscale mechanisms for high-pressure
  mechanochemistry: A phase field study}, Journal of Materials Science 53~(19)
  (2018) 13343--13363.
\newblock \href {https://doi.org/10.1007/S10853-018-2175-X}
  {\path{doi:10.1007/S10853-018-2175-X}}.

\bibitem{Javanbakht2016PhaseShear}
M.~Javanbakht, V.~I. Levitas, {Phase field simulations of plastic
  strain-induced phase transformations under high pressure and large shear},
  Phys. Rev. B 94 (2016) 214104.
\newblock \href {https://doi.org/10.1103/PhysRevB.94.214104}
  {\path{doi:10.1103/PhysRevB.94.214104}}.

\bibitem{Levitas2018Scale-FreeMicrostructure}
V.~I. Levitas, S.~E. Esfahani, I.~Ghamarian, {Scale-Free Modeling of Coupled
  Evolution of Discrete Dislocation Bands and Multivariant Martensitic
  Microstructure}, Physical Review Letters 121~(20) (2018) 205701.
\newblock \href
  {https://doi.org/10.1103/PHYSREVLETT.121.205701/FIGURES/4/MEDIUM}
  {\path{doi:10.1103/PHYSREVLETT.121.205701/FIGURES/4/MEDIUM}}.

\bibitem{Esfahani-etal-2020}
S.~E. Esfahani, I.~Ghamarian, V.~I. Levitas, {Strain-induced multivariant
  martensitic transformations: A scale-independent simulation of interaction
  between localized shear bands and microstructure}, Acta Materialia (2020)
  430--443.

\bibitem{Levitas_2017}
V.~I. Levitas, {Elastic model for stress tensor-induced martensitic
  transformation and lattice instability in silicon under large strains},
  Materials Research Letters 5~(8) (2017) 554--561.
\newblock \href {https://doi.org/10.1080/21663831.2017.1362054}
  {\path{doi:10.1080/21663831.2017.1362054}}.

\bibitem{Hill1984OnStrain}
R.~Hill, {On Macroscopic Effects Of Heterogeneity In Elastoplastic Media At
  Finite Strain}, Mathematical Proceedings of the Cambridge Philosophical
  Society 95~(3) (1984) 481--494.
\newblock \href {https://doi.org/10.1017/S0305004100061818}
  {\path{doi:10.1017/S0305004100061818}}.

\bibitem{Levitas1996SomeSurfaces}
V.~I. Levitas, Some relations for finite inelastic deformation of
  microheterogeneous materials with moving discontinuity surfaces, in:
  A.~Pineau, A.~Zaoui (Eds.), IUTAM Symposium on Micromechanics of Plasticity
  and Damage of Multiphase Materials, Springer Netherlands, Dordrecht, 1996,
  pp. 313--320.

\bibitem{Petryk1998MacroscopicTransformation}
H.~Petryk, {Macroscopic rate-variables in solids undergoing phase
  transformation}, Journal of the Mechanics and Physics of Solids 46~(5) (1998)
  873--894.
\newblock \href {https://doi.org/10.1016/S0022-5096(97)00099-9}
  {\path{doi:10.1016/S0022-5096(97)00099-9}}.

\bibitem{Schall-etal-2008}
J.~D. Schall, G.~Gao, J.~A. Harrison, {Elastic constants of silicon materials
  calculated as a function of temperature using a parametrization of the
  second-generation reactive empirical bond-order potential}, Physical Review B
  77~(11) (2008) 115209.

\bibitem{Bangerth2007Deal.IIALibrary}
W.~Bangerth, R.~Hartmann, G.~Kanschat, {deal.IIA general-purpose
  object-oriented finite element library}, ACM Transactions on Mathematical
  Software (TOMS) 33~(4) (8 2007).
\newblock \href {https://doi.org/10.1145/1268776.1268779}
  {\path{doi:10.1145/1268776.1268779}}.

\bibitem{groeber2014dream}
M.~A. Groeber, M.~A. Jackson, {DREAM. 3D: A digital representation environment
  for the analysis of microstructure in 3D}, Integrating Materials and
  Manufacturing Innovation 3~(1) (2014) 56--72.

\bibitem{zeng2020origin}
Z.~Zeng, Q.~Zeng, M.~Ge, B.~Chen, H.~Lou, X.~Chen, J.~Yan, W.~Yang, H.-k. Mao,
  D.~Yang, et~al., {Origin of plasticity in nanostructured silicon}, Physical
  Review Letters 124~(18) (2020) 185701.

\bibitem{xuan2020pressure}
Y.~Xuan, L.~Tan, B.~Cheng, F.~Zhang, X.~Chen, M.~Ge, Q.~Zeng, Z.~Zeng,
  {Pressure-induced phase transitions in nanostructured silicon}, The Journal
  of Physical Chemistry C 124~(49) (2020) 27089--27096.

\bibitem{tolbert1996pressure}
S.~H. Tolbert, A.~B. Herhold, L.~E. Brus, A.~Alivisatos, {Pressure-induced
  structural transformations in Si nanocrystals: surface and shape effects},
  Physical Review Letters 76~(23) (1996) 4384.

\end{thebibliography}

\end{document}